\documentclass{article}

\usepackage{multiequations}
\usepackage{graphicx}
\usepackage{xcolor}
\usepackage{amsfonts}
\usepackage{amsmath}
\usepackage{amssymb}
\usepackage{empheq} 
\usepackage[hidelinks]{hyperref}
\usepackage[round]{natbib}
\usepackage{enumerate}  
\usepackage[loadonly]{enumitem}  
\usepackage{bbm}

\usepackage{geometry}

\newcommand*{\be}{\begin{equation}}
\newcommand*{\ee}{\end{equation}}
\newcommand*{\bea}{\begin{eqnarray}}
\newcommand*{\eea}{\end{eqnarray}}
\newcommand*{\bal}{\begin{align}}
\newcommand*{\eal}{\end{align}}
\newcommand*{\bme}{\begin{multiequations}}
\newcommand*{\eme}{\end{multiequations}}

\newcommand{\defi}{\stackrel{\triangle}{=}}
\newcommand{\dv}{\bnabla\! \bcdot\!}

\newcommand{\vq}{Var(\varTheta)}

\newcommand{\E}{\mathbb{E}(}
\newcommand{\dfi}{\partial_{i}}
\newcommand{\dfj}{\partial_{j}}
\newcommand{\id}{\mbs{\mathbb{I}}_d}
\providecommand\bnabla{\boldsymbol{\nabla}}
\providecommand\bcdot{\boldsymbol{\cdot}}

\renewcommand*{\Omega}{\varOmega}
\renewcommand*{\Sigma}{\varSigma}

\newsavebox{\astrutbox}
\sbox{\astrutbox}{\rule[-5pt]{0pt}{20pt}}

\newcommand{\R}{\mathbb{R}}

\def\squarebox#1{\hbox to #1{\hfill\vbox to #1{\vfill}}}
\newcommand{\reel}{\mathbb R}

\newcommand{\Dt}{\mathbb{D}}

\newcommand{\defin}{\stackrel{\scriptscriptstyle\triangle}{=}}

\newcommand{\w}{\boldsymbol{w}}

\newcommand{\B}{\boldsymbol{B}}
\newcommand{\bsigma}{\boldsymbol{\sigma}}

\newcommand{\xx}{\boldsymbol{x}}

\newcommand{\XX}{\boldsymbol{X}}
\newcommand{\yy}{\boldsymbol{y}}
\newcommand{\zz}{\boldsymbol{z}}

\newcommand{\YY}{\boldsymbol{Y}}
\newcommand{\nab}{\boldsymbol{\nabla}}

\newcommand{\car}{1\!\mathsf{I}}

\newcommand{\transp}{^{\scriptscriptstyle T}}
\newcommand{\tr}{\mathrm{tr}}

\newcommand{\Exp}{\mathbb{E}}

\newcommand{\dif}{{\mathrm{d}}}
\newcommand{\mbs}[1]{\ensuremath{\boldsymbol{#1}}}
\newcommand{\mbf}[1]{\ensuremath{\boldsymbol{#1}}}
\newcommand{\dBt}{\dif  \boldsymbol{B}_t}

\begin{document}

%

\title{Geophysical flows under location uncertainty, Part I\\ Random transport
and general models}


\author{V. Resseguier${\dag\ddag}$$^{\ast}$\thanks{$^\ast$Corresponding author. Email: valentin.resseguier@inria.fr
\vspace{6pt}},
 E. M\'emin${\dag}$
and B. Chapron${\ddag}$
\\\vspace{6pt}  ${\dag}
$Inria/IRMAR/irstea, Fluminance group,\\ Campus universitaire de Beaulieu, Rennes Cedex 35042, France\\ ${\ddag}
$Ifremer, LOPS, Pointe du Diable, Plouzan\'e 29280, France
}

\maketitle

\begin{abstract}
A stochastic flow representation is considered with the Eulerian velocity decomposed between a smooth large scale component and a rough small-scale turbulent component. The latter is specified as a random field uncorrelated in time. Subsequently, the material derivative is modified and leads to a stochastic version of the material derivative to include a drift correction, an inhomogeneous and anisotropic diffusion, and a multiplicative noise. As derived, this stochastic transport exhibits a remarkable energy conservation property for any realizations. As demonstrated, this pivotal operator further provides elegant means to derive stochastic formulations of classical representations of geophysical flow dynamics.


\end{abstract}

\section{Introduction}
Despite the increasing power of computational resources and the availability of high quality observations, a precise description of geophysical flows over their whole dynamical scales is today completely beyond reach. Challenges appear as unlimited as the variety of dynamics and boundary conditions with their broad range of spatial and temporal scales across the globe. To face these challenges, numerous efforts are taking place to build an ever-increasing quality, quantity, duration and integration of all observations, in situ and satellite. In parallel, simulation capabilities largely improved, {\rm i.e.}, analysis can now be routinely carried out to more precisely characterize the variability in the global ocean, at scales of ten to hundreds of kilometers and one to hundreds of days. Yet, for these ocean models, the unresolved small scales and associated fluxes are always accounted for by simple mathematical models, {\rm i.e.} parameterizations.

Although the development of more efficient sub-grid representations remains a very active research area, the possible separation between relatively low-frequency, large scale patterns and transient, small-scale fluctuations, strongly invites to consider stochastic representations of the geophysical dynamics \citep[e.g.][]{Hasselmann76,allen2002towards,penland2003stochastic,berner2011model,Franzke15}. As derived, such developments are meant to better describe the system's variability, especially including a mean drift , called ``bolus" velocity \citep{Gent90} or skew-diffusion \citep{Nakamura01,Vallis} in oceanography, and noise-induced velocity in climate sciences.

In that context, several different strategies have been proposed \citep{Franzke15}. Among them, techniques motivated by physics have been devised. Those schemes aim to overcome a bad representation of the small scale forcing and of their interactions with the large scale processes. Two of such schemes have been carried out at ECMWF. The first one, the stochastic perturbation of the physical tendencies -- SPPT -- \citep{Buizza99} implements a multiplicative random perturbation of parameterized physical tendencies. The random variables involved are correlated in space and time, and their characteristics set from fine grid simulations. The second one, the stochastic kinetic-energy backscatter -- SKEB -- \citep{Shutts05} introduces a perturbation of the stream function and potential temperature. This scheme is based on earlier works on energy backscattering modelling through the introduction of random variables \citep{Mason92}. Numerous works  showed a beneficial impact of the injected randomness on weather and climate forecasts mean and variability (see \citep{Berner15} and references therein) or in oceanography \citep{Brankart13,mana2014toward}. However, the amplitude of the perturbations to apply is difficult to specify. The non-conservative and the variance-creating nature of those schemes is also problematic in that prospect. A too large amplitude, while increasing significantly the ensemble spread, may lead to unstable schemes for simulations that go beyond short-term forecast applications. A balance between the large-scale sub-grid diffusive tensor and the noise amplitude must thus be found to stabilize the system. 

Also based on a separation of the state variables between slow and fast components, a mathematical framework -- refereed to as MTV algorithms -- has been proposed to derive stochastic reduced-order dynamical systems for weather and climate modelling \citep{Franzke05,Franzke06,Majda99,Majda01,Majda03}. Considering a linear stochastic equation to describe the fast modes, derivations have been rigorously studied \citep{Gottwald13,Melbourne11,Pavliotis08}. As demonstrated, the continuous fast dynamics converges in continuous time towards a Stratonovich noise, leading to a diffusion term when expressed in a corresponding Ito stochastic integral form.

As well, stochastic superparametrization assumes a scale separation \citep{Grooms13,Grooms14}. The point approximation and Reynolds decompositions replace homogenization techniques. As for MTV methods, the small-scale evolution law is linearized and corrected with the introduction of noise and damping terms. The second order moments of the solution are then known analytically and can feed the sub-grid tensors expression of the mean deterministic large-scale evolution law.
For such developments, the direct use of the Reynolds decomposition implicitly assumes that small-scale components are differentiable. This theoretically prevents the use of Langevin type equations for the small-scale evolution. Furthermore, in such a derivation, each scalar evolution law involves a different sub-grid tensor. Similarly to the definition of eddy viscosity and diffusivity models for Large-Eddy simulation, the noise expression of most stochastic fluid dynamic models
are hardly inferred from physics. So, instantaneous diffusion and randomness may not be consistently related; even though some careful parametrizations of stationary energy fluxes couple them \citep{Grooms13,sapsis2013blending,Grooms14,sapsis2013statistically}.

To overcome these difficulties, we propose to dwell on a different strategy. As previously initiated \citep{Memin14}, the large-scale dynamics is not prescribed from a deterministic representation of the system's dynamics. Instead, a random variable, referred to as location uncertainty, is added to the Lagrangian expression of the flow. The resulting Eulerian expression then provides stochastic extensions of the material derivative and of the Reynolds transport theorem. An explicit expression of a noise-induced drift is further obtained. As also derived, a sub-grid stress tensor, describing the small-scale action on the large scales, does not resort to the usual Boussinesq eddy viscosity assumption, and further, consistently appears throughout all the conservation equations of the system. Moreover, the advection by the unresolved velocity acts as a random forcing. As such, this framework provides a direct way to link the resulting material transport and the underlying dynamics. The well-posedness of these equations has been studied by \cite{mikulevicius2004stochastic} and \cite{flandoli2011interaction}.
Recently, \cite{Holm2015} derived similar evolution laws from the inviscid and adiabatic framework of Lagrangian mechanics. Compared to models under location uncertainty, the stochastic transport of scalars is identical. However, the momentum evolution of \cite{Holm2015} involves an additional term which imposes the helicity conservation but may increase the kinetic energy.

Starting with the description of the transport under location uncertainty (section 2), developments are then carried out to explore this stochastic framework for different classical geophysical dynamical models (section 3).

\section{Transport under location uncertainty}
\label{Transport under location uncertainty}

\subsection{A 2-scale random advection-diffusion description}
\label{Informal description}

As often stated, ocean and atmospheric dynamics can be assumed to be split into two contributions with very distinct correlation times. This assumption can especially hold for the top layer of the ocean. For example, the larger ocean geostrophic component generally varies on much slower time scales than motions at smaller spatial scales. From an observational perspective, current generation satellite altimeter instruments are capable of resolving only the largest eddy scales, and the measurements can depend sensitively on the local kinetic energy spectrum of the unresolved flow \citep{poje2010resolution,keating2011diagnosing}. Satellite observations of the upper-ocean velocity field at higher resolution can also be obtained \citep[e.g.][]{chapron2005direct} but are certainly too sparse and possibly noisy. 

Accordingly, without loss of generality, observations of an instantaneous Eulerian velocity field are likely coarse-grained in time, and can be interpreted under a 2-scale framework. As such, the instantaneous Eulerian velocity is decomposed between a well resolved smooth component, denoted $\mbs w$, continuous in time, and a rough small-scale one, rapidly decorrelating in time. This badly-resolved contribution, expressed as $\bsigma \dot{\mbs B}$, is then assumed Gaussian, correlated in space, but uncorrelated in time. This contribution can be inhomogeneous and anisotropic in space. Due to the irregularity of the flow, the transport of a conserved quantity, $\varTheta$, by the whole velocity, defined as 
\bea
 \varTheta 
 (\XX_{t + \Delta t},t + \Delta t) 
 &=&
 \varTheta (\XX_{t},t)  
\eea
corresponds to a random mapping. 
In this setup the large-scale velocity possibly depends on the past  history of the small-scale component. This latter being white in time, the two components are uncorrelated.
Hence, the above conservation shall lead to a classical advection-diffusion evolution, with the introduction of an inhomogeneous and anisotropic diffusion coefficient matrix, $\mbs a$, solely defined by the one-point one-time covariance of the unresolved displacement per unit of time:
\bea 
\mbs a = 
\frac{
\Exp \left \{ \bsigma \dBt \left( \bsigma \dBt  \right)\transp  \right \} 
}{\dif t}
.
\label{balance}
\eea
The inhomogeneous structure of the small-scale variance motions shall  
create inhomogeneous spreading rates. More agitated fluid parcels  
spread faster than those over quiescent regions. Overall, the latter  can be seen as ``attracting'' the large-scale gradients. This effect leads to invoke a  drift correction, anti-correlated with the variance gradient, or, in a multi-dimensional point of view, anti-correlated with the covariance matrix divergence. Accordingly, the random advection under a 2-scale description can be expected to be expressed as:
\bea
\partial_t \varTheta +
\underbrace{
{\w}^\star\bcdot \nab \varTheta
}_{\text{Corrected advection}}
=
\underbrace{
\nab\bcdot \left ( \tfrac{1}{2} \mbs a  \nab \varTheta \right )
}_{\text{Diffusion}}
-
\underbrace{
 \bsigma \dot{\mbs B} \bcdot \nab \varTheta
}_{\text{Random forcing}},
\label{heuristic transport}
\eea
with a modified velocity given by
\bea
\w^\star = \w   - \tfrac{1}{2} ( \nab\bcdot \mbs a)\transp
+ \bsigma (\dv \bsigma)\transp
.
\eea
We note the conserved quantity is diffused by the small-scale random velocity. 
The random forcing expresses the advection by the unresolved velocity $\bsigma \dot{\mbs B}=\bsigma{\dBt}/{\dif t}$, and continuously backscatters random energy to the system. Because of this white-noise forcing term, the Eulerian conservation equation \eqref{heuristic transport} (that will be formally expressed in the following sections) intrinsically concerns  a random non-differentiable tracer. 
Finally, the conserved quantity is also advected by an ``effective" velocity, $\w^\star$, taking into account the possible spatial variation of the small-scale velocity variance,   as well as the possible divergence of this velocity component.

Considering the unresolved velocity and this effective drift, $\w^\star$, divergent-free, we shall see that this 2-scale development establishes an exact balance between the amount of diffusion and the random forcing. Subsequently, essential properties related to energy conservation and mean/variance tracer evolution directly result from this balance.

\subsection{Uncertainty formalism}
\label{The used stochastic model}

In a Lagrangian stochastic form, the infinitesimal displacement associated with a particle trajectory $\XX_t$ is:
\begin{eqnarray}
\label{particle_dX}
\dif\XX_t &=& \w(\XX_t,t) \dif t+ \bsigma(\XX_t,t) \dif\B_t.
\end{eqnarray}
Formally, this is defined over the fluid domain, $\Omega$, from a $d$-dimensional Brownian function $\B_t$. Such a function can be interpreted as a white noise process in space and a Brownian process in time\footnote{Formally it is a cylindrical $I_d$-Wiener process (see \cite{DaPrato} and \cite{Prevot07} for more information on infinite dimensional Wiener process and cylindrical $I_d$-Wiener process).}. The time derivative of the Brownian function, in a distribution sense, is denoted  $\bsigma \dot{\mbs B} =\bsigma  {\dBt}/{\dif t}$, and is a white noise distribution. The spatial correlations of the flow uncertainty are specified through the diffusion operator $\bsigma(.,t)$, defined for any vectorial function, $\mbs f $,  through the matrix kernel $\breve{\bsigma} (.,.,t)$:
 \begin{equation}
\bsigma(\xx,t)\mbs f \defin \int_\Omega \breve{\bsigma}(\xx,\zz,t)  \mbs f (\zz,t) \dif\zz.
\end{equation}
This quantity is assumed to have a finite norm\footnote{More precisely, the operator $\bsigma$ is assumed to be Hilbert-Schmidt.}
and to have a null boundary condition on the domain frontier\footnote{Note that periodic boundary conditions can also be envisaged.}.
The resulting $d$-dimensional random field, $\bsigma(\xx,t) \dBt$, is a centered vectorial Gaussian function, correlated in space and uncorrelated in time with covariance tensor:
\bea
\mbs {Cov} (\xx,\yy,t,t')
 &\defi&
\Exp \left \{
\left (\bsigma(\xx,t) \dif\B_t \right ) \left ( \bsigma(\yy,t') \dif\B_{t'} \right ) \transp 
\right \}
,\\
&=&
\int_\Omega \breve \bsigma(\xx,\zz,t) \breve\bsigma \transp (\yy,\zz,t)\dif\zz
\ \delta(t-t')\dif t.
\eea
For sake of thoroughness,  the uncertainty random field has a (mean) bounded norm\footnote{
This norm is finite since $\bsigma$ is Hilbert-Schmidt, ensuring the boundness of the trace of operator $Q$ -- defined by the kernel $(\xx,\yy) \mapsto \bsigma(\xx,t) \bsigma \transp (\yy,t)$ --, and $ \forall t \leqslant T < \infty, \ 
\Exp \|\int_0^{t}\bsigma \dif\B_{t'}\|^2_{L^2(\Omega)}
=
\int_0^{t}\int_\Omega \| \breve \bsigma (\bullet, \zz) \|^2_{L^2(\Omega)}\dif \zz \dif t'
=
\int_0^{t} \| \bsigma \|^2_{HS,L^2(\Omega)} \dif t'
=
\int_0^{t}
tr(\mbs Q) \dif t'
<
\infty$, where the index HS refers to the Hilbert-Schmidt norm.}: $\Exp \| \int_0^t \bsigma \dif\B_{t'}\|^2_{L^2(\Omega)}  < \infty$ for any bounded time $t\leqslant T<\infty$.
Hereafter, the diagonal of the covariance tensor, $\mbs a$, will be referred to as the variance tensor:
\begin{eqnarray*}
\mbs a(\xx,t)\delta(t-t')\dif t 
= \mbs {Cov}(\xx,\xx,t,t') .
\end{eqnarray*}
By definition, it is a symmetric positive definite matrix at all spatial points, $\xx$. This quantity, also denoted $\bsigma \bsigma \transp$, corresponds to the time derivative of the so-called quadratic variation process:
\begin{eqnarray*}
\bsigma \bsigma \transp 
\defi 
\mbs a
=
\partial_t
\left < \int_0^t  \bsigma \dif \B_{s}, 
\left( \int_0^t  \bsigma \dif \B_{r} \right) \transp
\right >.
\end{eqnarray*}
with $\left<f,g\right>$ to stand for the quadratic cross-variation process of $f$ and $g$ (see Appendix \ref{QuadVar}).

Given this strictly defined flow, the corresponding material derivative expression of a given quantity can be introduced. 
\subsection{Material derivative}
To derive the expression of the material derivative $ \rm D_t \varTheta \defi \left ( \dif \left ( \varTheta \left( \XX_t,t \right ) \right ) \right )_{|_{\XX_t = x}} $, also quoted as the Ito-Wentzell derivative or generalized Ito derivative in a stochastic flow context \citep[theorem 3.2.2]{Kunita}, let us introduce 
an operator, hereafter referred to as the stochastic transport operator:
\bea
{ \Dt}_t \varTheta \ 
&\defi&
\underbrace{
\dif_t \varTheta
}_{
\substack{
 \defi \  \varTheta(\xx,t+\dif t) - \varTheta(\xx,t) \\
\text{Time increment}
}
}
+ 
\underbrace{
\left ({\w}^\star\dif t + \bsigma \dif\B_t \right)\bcdot \nab \varTheta
}_{\text{Advection}}
 - 
\underbrace{
\nab\bcdot \left ( \tfrac{1}{2} \mbs a  \nab \varTheta \right )
}_{\text{Diffusion}}
\dif t 
\label{Mder}
\eea 
This operator corresponds to a strict formulation of \eqref{heuristic transport}. More specifically, it involves a time increment term $\dif_t \varTheta$  instead of a partial time derivative as $\varTheta$ is non differentiable. Contrary to the material derivative, the transport operator has an explicit expression (equation \eqref{Mder}). However, the material derivative is explicitly related to the transport operator (see proof in Appendix \ref{link Material-Der})

\bea 
\label{link DD and material deriv}
\left\{
\begin{array}{r c l}
{ \Dt}_t \varTheta &= &
f_1 \dif t +  \mbs h_1 \transp \dif\B_t, \label{Dt-eq1}\\
\rm D_t \varTheta &=& 
f_2 \dif t +  \mbs h_2 \transp \dif\B_t,
  \end{array}
  \right.
\Longleftrightarrow 
\left\{
\begin{array}{r c l}
f_2 &=& 
f_1  
+ \tr\bigl(\left ( \bsigma \transp \bnabla \right )  \mbs h_1 \transp \bigr),\\
\mbs h_1 &=& 
\mbs h_2 .
  \end{array}
    \right.
  \eea 
Note, the material derivative, $\rm D_t$, has a clear physical meaning but no explicit expression whereas the explicit expression of the transport operator offers elegant means to derive stochastic Eulerian evolution laws. Most often both operators coincide and can interchangeably be  used. 
As a matter of fact, in most cases, we deal with null Brownian function $\mbs h_1$ in \eqref{link DD and material deriv}.
This corresponds, for instance, either to the transport of a scalar $\Dt_t \varTheta=0$ or to the conservation of an extensive property $\left( \int_{{\cal V} (t)} q\right)$ when the unresolved velocity component is solenoidal ($\dv \bsigma \dBt= 0$), which leads, as we will see it, to $\rm D_t q = -\dv \w^* q  \dif t$ (\eqref{th_transport}).
 In such a case, it is straightforward to infer from the system \eqref{link DD and material deriv}, that ${\Dt}_t$ and $\rm D_t$ coincide. For this precise case, those operators lead to
 \bea 
  \Dt_t \varTheta  (\XX_t,t) 
 = 
 \rm D_t \varTheta   (\XX_t,t) 
=
\dif\left( \varTheta (\XX_t,t) \right) 
= 
  f_1 (\XX_t,t)\dif t .
 \eea 
Going back to the Eulerian space, the classical calculus rules apply to operator $\Dt_t$, e.g. the product rule
\begin{equation}
\Dt_t(fg) (\xx,t) = \left( \Dt_t f \ g+ f \ \Dt_t g \right)(\xx,t),
\label{product-rule}
\end{equation} and the chain rule:
\begin{equation}
\Dt_t\bigl( \varphi\circ f \bigr) (\xx,t)= \Dt_t f  (\xx,t) (\varphi'\circ f)(\xx,t).
\label{chain-rule}
\end{equation}    
Given these properties, an expression for the stochastic advection of a scalar quantity can be derived.
\subsection{Scalar advection}
\label{Passive scalar advection}

The advection of a scalar $\varTheta$ thus reads:
\bea 
\label{eq Scalar advection}
\Dt_t \varTheta  = \rm D_t \varTheta = 0.
\eea 
To analyze this stochastic transport equation, let us first consider that the effective drift and the unresolved velocity are both divergence-free. As shown later, these conditions ensure an isochoric stochastic flow (see \eqref{eq_incomp_sto}).
With these conditions, the stochastic transport equation  exhibits remarkable conservation properties. 

 \subsubsection{Energy conservation}
 
 From (\ref{Mder}-\ref{eq Scalar advection}) and Ito lemma, the scalar energy evolution is given by:
 \begin{align}
 \dif \int_\Omega \tfrac{1}{2} \varTheta^2
 & =\int_\Omega \left(  \varTheta \dif_t \varTheta + \tfrac{1}{2} \dif_t \langle \varTheta,\varTheta\rangle \right),\\
 &=
 -\int_\Omega
   \tfrac{1}{2} \left( \w^*\dif t + \bsigma \dif\B_t \right) \bcdot \nab \left(\varTheta^2\right)
+ 
\underbrace{
 \int_\Omega \varTheta  \nab \bcdot \left(  \tfrac{1}{2} \mbs a  \nab \varTheta \right) \dif t 
}_{\text{Loss by diffusion}}
+
\underbrace{
 \int_\Omega
 \tfrac{1}{2}  \left( \nab \varTheta\right) \transp \mbs a  \nab \varTheta \dif t
}_{\text{Energy intake from noise}}
.
 \label{decomposition of the energy}
 \end{align}
For suitable boundary conditions, the two last terms cancel out after integration by part. The diffused energy is thus exactly compensated by the energy brought by the noise. With divergent-free conditions for  $\w^\star$ and $\bsigma$, another integration by part gives
 \bea
 \dif \int_\Omega \tfrac{1}{2} \varTheta^2
 &=&
\int_\Omega \tfrac{1}{2} \nab \bcdot \left( \w^*\dif t + \bsigma \dif\B_t \right)  \varTheta^2
 =0.
  \label{E-cons}
 \eea
The energy is thus conserved for all scalar random realizations. 
The expectation of the energy -- the energy (ensemble) mean -- is therefore also conserved. Moreover, from the decomposition
$\varTheta = \Exp(\varTheta) + \bigl(\varTheta -\Exp(\varTheta)\bigr)$ into the mean and the random anomaly component, we obtain a partition of this constant energy mean:
 \begin{equation}
 0=\frac{\dif}{\dif t} \Exp \|\varTheta\|^2_{{\cal L}^2(\Omega)}=  \frac{\dif}{\dif t} \|\Exp(\varTheta)\|^2_{{\cal L}^2(\Omega)} + \frac{\dif}{\dif t} \int_\Omega Var(\varTheta) .
 \label{Eq-Exp-var}
 \end{equation}
A decrease of the mean energy -- the energy of the (ensemble) mean --  is always associated with an (ensemble) variance increase. Similar energy mean budgets have recently been discussed by several authors. \cite{majda2015statistical} refers to this energy mean as the statistical energy. The author derives the evolution law of this energy by adding the evolution equations of the mean energy and of the integrated variance, whereas our energy budget is obtained by evaluating the mean of the evolution law of the total energy, $ \|\varTheta\|^2_{{\cal L}^2(\Omega)}$. However, \cite{majda2015statistical} does not specify the random forcing. This is why the latter does not {\em a priori} balance the turbulent diffusion.
\cite{Farrell14} also studied the energy mean of stochastic fluid dynamics systems especially under quasi-linear approximations and with an additive Gaussian forcing.

By the chain rule, all the tracer moments are also conserved:
\begin{equation} 
{\Dt}_t \varTheta^p = p \ \varTheta^{p-1} {\Dt}_t \varTheta =0.
\end{equation}
Yet, the energy of statistical moments are in general not conserved, as detailed in the following section.

 \subsubsection{Mean and variance fields of a passive scalar}

Consider now that the expectation corresponds to a conditional expectation given the effective drift. This applies to passive scalar transport for which the drift does not depend on the tracer. Terms in $\dBt$ have zero-mean, and the mean passive scalar evolution can be immediately derived taking the conditional expectation of the stochastic transport:
 \begin{equation}
 \label{mean tracer}
\partial_t \Exp (\varTheta)
+
\underbrace{
 \w^\star \bcdot \nab \Exp (\varTheta) 
  }_{\text{Advection}}
  = 
\underbrace{
  \nab\bcdot \left(\tfrac{1}{2}\mbs a \nab \Exp(\varTheta)\right)
  }_{\text{Diffusion}}
  .
  \end{equation}
Since $\mbs w^*$ is divergent-free, it has no influence on the energy budget. The mean field energy decreases with time due to diffusion. As for the variance, its evolution equation, derived in Appendix \ref{variance tracer proof}, reads:
    \begin{equation}
    \label{variance tracer}
\partial_t Var (\varTheta) 
+
\underbrace{
  \w^\star \bcdot \nab Var (\varTheta)
  }_{\text{Advection}}
=
\underbrace{
\nab\bcdot
\left(\tfrac{1}{2}\mbs a \nab Var(\varTheta)\right) 
  }_{\text{Diffusion}}
  +
\underbrace{
\left(\nab\Exp(\varTheta) \right )\transp \mbs a \nab \Exp(\varTheta)
  }_{\text{Variance intake}}
  .
  \end{equation}
This is also an advection-diffusion equation, with an additional source term. Integrating this equation on the whole domain, with the divergent-free condition, and considering the divergence form of the first right-hand term, we obtain
  \begin{equation}
\frac{\dif}{\dif t}\int_\Omega Var (\varTheta)
=
\int_\Omega \left(\nab\Exp(\varTheta) \right )\transp \mbs a \nab \Exp(\varTheta)
\geqslant 0.
  \end{equation}
It shows that the stochastic transport of a passive scalar creates variance. The dissipation that occurs in the mean-field energy equation is exactly compensated by a variance increase. This mechanism is very relevant for ensemble-based simulations. The uncertainty modeling directly incorporates a large-scale dissipating sub-grid tensor, and further encompasses a variance increase mechanism to balance the total energy dissipation. Such a mechanism is absent in ensemble-based data assimilation development
\citep{berner2011model,gottwald2013role,Snyder15}. An artificial inflation of the ensemble variance is  usually required in consequence to avoid filter divergence \citep{Anderson99}. 

 \subsubsection{Active tracers}

%

For the more general case of an active tracer, the velocity depends on the tracer distribution, additional energy transfers occurs between the mean and the random tracer components \citep{sapsis2013attractor,sapsis2013blending,sapsis2013statistically,ueckermann2013numerical,majda2015statistical}. Though a complete analytical description is involved, these energy transfers are mainly due to the nonlinearity of the flow dynamics, and are hence more familiar. The models under location uncertainty involve both types of interactions: the ``usual'' nonlinear interactions and the random energy transfers previously described. As such, these two energy fluxes analyzes are complementary. In deterministic turbulent dynamics with random initial conditions, energy is drained from the mean tracer toward several modes  ({\em e.g.} Fourier modes) of the tracer random component, and is backscattered from other modes. The energy fluxes toward (from) random modes increases (decreases) the variance. In the case of the deterministic Navier-Stokes equations, \cite{sapsis2013attractor} analytically expressed the integrated variance. The molecular or turbulent diffusion decreases the variance whereas the mean velocity may increases or decreases the random energy, by triad interactions. The modes receiving energy become unstable, whereas those giving energy are over-stabilized \citep{sapsis2013blending}.
In ensemble data assimilation of large-scale geophysical flows, the solution is defined by a manifold sampled by a small ensemble of realizations. Those stabilizations and destabilizations are the reason for the alignment of ensembles along unstable directions \citep{trevisan2004assimilation,ng2011role}. It can lead to filter divergence \citep{gottwald2013role,bocquet2016degenerate}.
In the absence of  any modes truncation, the nonlinear interactions redistribute the energy between those modes. Otherwise, the missing energy fluxes can be parametrized with additional random terms \citep{sapsis2013blending,sapsis2013statistically}.

To further describe the energy exchanges involved in the dynamics under location uncertainty of active tracers,
we introduce the decomposition $\varTheta=\widetilde \varTheta + \varTheta'$ in terms of a slow component $\widetilde \varTheta$ and a highly oscillating component $\varTheta'$. The first one is time-differentiable whereas the second is only continuous with respect to time. Both components are random. 
This decomposition, the so-called semi-martingale decomposition, is unique \citep{Kunita}. For each component, the following coupled system of transport equations is:
\begin{align}
&\partial_t \widetilde \varTheta
+ \w^\star \bcdot \nab \varTheta
= 
\nab\bcdot \left(\tfrac{1}{2}\mbs a \nab \varTheta\right), 
\label{eq-dq-tilde}\\
&\dif_t \varTheta' + \bsigma\dif\B_t \bcdot \nab \varTheta =0.
\label{eq-dq'}
\end{align}
At the initial time, the first component is deterministic (given the initial conditions) and the second one is zero. The large-scale component becomes random through the oscillating component, which is characterized  by a gradually increasing energy along time:  
\begin{equation}
 \Exp \|\varTheta'\|^2_{L^2(\Omega)}
= \Exp \int_\Omega  \langle \varTheta',\varTheta'\rangle 
= \Exp \int_0^t \int_\Omega   \left(\nab \varTheta \right )\transp \mbs a \nab \varTheta \; \dif t
\geqslant 0
.
\end{equation}
Note, the expectation is taken with respect to the law of the Brownian path. The energy mean of the non-differentiable component $\varTheta'$ is the mean of the energy intake provided by the noise \eqref{decomposition of the energy}. The same amount of energy mean is removed from the system by the diffusion \eqref{decomposition of the energy}. Once diffused, this energy is fed back to the small-scale tracer $\varTheta'$, the white noise velocity acting here as an energy bridge. Such an energy redistribution is a main issue in sub-grid modeling. 
Indeed, as explained above,  large-scale flow simulations often miss to capture the energy fluxes between the mean and the random components but also the energy redistribution from the unstable modes to the stable modes.
Note that, even though the two components are orthogonal as functions of time (in a precise sense), they are not, in general, as functions of space: $\int_{\Omega}\widetilde{\varTheta}\varTheta'\neq 0$. In particular, it can be shown that those two components are indeed anti-correlated when the tracer is passive.

\subsubsection{The homogeneous case and the Kraichnan model}

A divergent-free isotropic random field for the small-scale velocity component corresponds to the Kraichnan model \citep{Kraichnan68,kraichnan1994anomalous,Gawedzki95,Majda-Kramer}. The variance tensor, $\mbs a$, becomes a constant diagonal matrix $\frac 1 d tr(\mbs a) \id$, where $d$ stands for the dimension of the spatial domain $\Omega$. The tracer evolution now involves a Laplacian diffusion
\begin{align}
\dif_t \varTheta +  \bigl(\w \dif t + \bsigma \dif \B_t\bigr )  \bcdot \nab \varTheta
=  \frac{tr(\mbs a)}{2d}   \Delta \varTheta \dif t.
\end{align}

Additionally, the original Kraichnan model considers a small molecular diffusion, $\nu$, and an external Gaussian forcing, $ f \dif B_t'$, defined as an homogeneous random field uncorrelated in time and independent of the velocity component $\bsigma \dot{\mbs B}$ \citep{Gawedzki95}. 
In our framework, the Kraichnan model, which does not involve any large-scale drift term, reads:
\bea 
\dif_t \varTheta +   \bsigma \dif \B_t \bcdot \nab \varTheta
=  
\left( \nu + \frac{tr(\mbs a)}{2d}   \right) \Delta \varTheta \dif t 
+ f \dif B_t'.
\eea
As compared to the original model, this derivation directly identifies the eddy diffusivity contribution, only implicitly termed in the Kraichnan model \citep{Gawedzki95,Majda-Kramer}. 
This usual formulation corresponds to the Stratonovich notation.
The Ito calculus further offers means to infer the evolution of the tracer moments, \eqref{mean tracer} and \eqref{variance tracer}. The proposed development introduces an additional non-linearity through $\mbs w$ and possible non-uniform turbulence conditions. 

\subsection{Transport of extensive properties}

Hereafter, all fundamental conservation laws are formulated for extensive properties. 

\subsubsection{Stochastic Reynolds transport theorem}
\label{subsubsection Stochastic Reynolds transport theorem}

Similar to the deterministic case, the stochastic Reynolds transport theorem shall describe the time differential of a scalar function, $q(\xx,t)$, within a material volume, ${\cal V}(t)$, transported by the random flow (\ref{particle_dX}):
\begin{equation}
\label{th_transport}
\dif \int_{{\cal V}(t)} q
= 
\int_{{\cal V} (t)} 
\biggl[
\rm D_t q 
  + \nab \bcdot
  \left( \w^\star   \dif t + \bsigma \dBt \right )q
  +
 \dif  \left <
  \int_0^t D_{t'} q ,
  \int_0^t \dv \bsigma \dif \B_{t'}
   \right >
  \biggr].
\end{equation} 
This expression, rigorously derived in Appendix \ref{Reynolds}, was first introduced in a slightly different version by \cite{Memin14}. In most cases, the unresolved velocity component, $\bsigma \dot{\mbs B}$, is divergence-free and, the source of variations of the extensive property $\int_{{\cal V}(t)}q$ is time-differentiable, {\rm i.e.} with a differential of the form $
\dif \int_{{\cal V}(t)} q = \mathcal{F} \dif t
$. In such a case, for an arbitrary volume, the transport theorem takes the form $\rm D_t q = f \dif t$, and according to equation \eqref{link DD and material deriv} the material derivative can be replaced by the stochastic transport operator, $\Dt_t q$, to provide an intrinsic expression of this stochastic transport theorem.
\subsubsection{Jacobian}
Taking $q=1$ characterizes the volume variations through the flow Jacobian, $J$:
\begin{subequations}
\bea
\int_{{\cal V} (t_0)}\!\!\!\!\!\!\!\!\dif (J(\XX_t(\xx_0),t)) \dif \xx_0
&=&
\dif \!\!\int_{{\cal V} (t)} \!\!\!\!\!\!\! \dif \xx 
,\\
&=&
\int_{{\cal V} (t)} \!\!\!\!\!\!\!\!
\nab \bcdot  \left( \w^\star   \dif t + \bsigma \dBt \right )(\xx,t) \;\dif \xx 
,\\
&=&
\int_{{\cal V} (t_0)} 
\bigg [
J \nab\bcdot  \left( \w^\star   \dif t + \bsigma \dBt \right )
\bigg ]  (\XX_t(\xx_0),t)
 \; \dif \xx_0 .
\label{th_transport-formal}
\eea
\end{subequations}
Valid for an arbitrary initial volume ${\cal V}(t_0)$, it leads to a familiar form for the Lagrangian flow Jacobian evolution law:
\bea 
\rm D_t J  - J \nab\bcdot  \left( \w^\star   \dif t + \bsigma \dBt \right ) = 0.
\label{Jacobian eq}
\eea

\subsubsection{Incompressibility condition}

The Jacobian evolution \eqref{Jacobian eq} ensures a necessary and sufficient condition for the isochoric nature of the stochastic flow:
\begin{eqnarray}
\label{eq_incomp_sto}
\nab\bcdot \bsigma = 0
\text{ and }
\nab\bcdot \mbs w^*
= 0.
\end{eqnarray} 
If the large-scale flow component, $\w$, is solenoidal, this reduces to:
\begin{equation}
\label{incompressibility}
\nab\bcdot \bsigma= 0
\text{ and }
\nab\bcdot \w = \nab\bcdot \left( \nab\bcdot \mbs a \right) \transp=0.
\end{equation}
Note that for an isotropic unresolved velocity, the last condition is naturally satisfied, as this unresolved velocity component is associated with a constant variance tensor, $\mbs a$.
\subsection{Summary}

An additional Gaussian and time-uncorrelated velocity modifies the expression of the material derivative. In most cases, the resulting stochastic transport operator, $\Dt_t$, coincides with the material derivative, $\rm D_t$. Yet, possible differences between $\Dt_t$ and $\rm D_t$ have simple analytic expressions. This stochastic transport operator leads to an Eulerian expression of the tracer transport. As obtained, the tracer is forced by a multiplicative noise and mixed by an inhomogeneous and anisotropic diffusion. Moreover, the advection drift is possibly modified with a correction term related to the spatial variation of the small-scale velocity variance. The random forcing, the dissipation and the effective drift correction are all linked. Accordingly, the energy is conserved for each realization, as the tracer energy dissipated by the diffusion term is exactly compensated by the energy associated with the random velocity forcing. For a passive tracer, the evolution laws for the mean and variance precise these energy exchanges. The unresolved velocity transfers energy from the mean part of the tracer to its random part. For an active tracer, this velocity component bears energy from the whole tracer field to its random non-differentiable component. 



\section{Stochastic versions of geophysical flow models}
\label{section Stochastic versions of oceanic models}

The stochastic version of the Reynolds transport theorem provides us the flow Jacobian evolution law, as well as the rate of  change expression of any scalar quantity within a material volume. Together with the fundamental conservation laws of classical mechanics, it provides us a powerful tool to  derive in a systematic way stochastic flow models. Thanks to the bridge between the material derivative and the stochastic transport operator, this derivation closely follows the usual deterministic derivations.

All along the following development, the small-scale random flow component will be assume incompressible, {\rm i.e.} associated with a divergence-free diffusion tensor:
\bea 
\label{incompressiblity cond sigma}
\nab\bcdot \bsigma=0.
\eea 
This assumption remains realistic for the geophysical models considered in this study, and does not prevent the resolved velocity component (and therefore the whole field) to be compressible.


\subsection{Mass conservation}
Mass conservation for arbitrary volumes rules the stochastic transport of the fluid density, denoted $\rho$:
\begin{equation}
\label{Cont-eq}
\Dt_t \rho + \rho \nab\bcdot \w^* \dif t=0.
\end{equation} 
A suggested in \ref{subsubsection Stochastic Reynolds transport theorem}, the material derivative, $\rm D_t$, is now replaced by $\Dt_t$, defined by Eq. \eqref{Mder}. Indeed, the mass variation is zero and thus time-continuous, and the stochastic operator coincides with the material derivative.

\subsection{Active scalar conservation law}

The transport theorem \eqref{th_transport} applied to a quantity $\rho \varTheta$ describes the rate of change of the scalar $\varTheta$ and is generally balanced by a production/dissipation term, as:
\begin{equation}
\label{sto-scalar-cons-gen}
 \Dt_t (\rho\varTheta) +\rho \varTheta \nab\bcdot \w^* \dif t
= \rho{\cal F}_\varTheta(\varTheta) \dif t.
\end{equation}
Again, the stochastic transport operator, $\Dt_t$, is used instead of the material derivative, $\rm D_t$, since the source of variation $\int_0^t \left( \int_{{\cal V}(t)} \rho{\cal F}_\varTheta \right) \dif t$ of the extensive property, $\int_{{\cal V}(t)} \rho\varTheta$, is time-differentiable (integral in $\dif t$), as explained in \ref{subsubsection Stochastic Reynolds transport theorem}. 
Considering the product rule (\ref{product-rule}) and mass conservation \eqref{Cont-eq}, the transport  evolution model for the scalar writes: 
\begin{equation}
\label{sto-scalar-cons}
 \Dt_t\varTheta =  {\cal F}_\varTheta(\varTheta) \dif t.
\end{equation}

For a negligible production/dissipation term,
the scalar is conserved by the stochastic flow and follows properties highlighted in section \ref{Transport under location uncertainty} -- e.g. the energy conservation of each realization and the dissipation of the mean field.
As in the deterministic case, the 1$^{st}$ law of thermodynamics implies both temperature conservation ($\varTheta=T$) and conservation of the amount of substance -- e.g. the conservation of salinity ($\varTheta=S$):
\begin{subequations}
\bea 
\label{transportTemp}
\Dt_t T={\cal F}_T(T) \dif t,\\
\label{transportSalinity}
\Dt_t S ={\cal F}_S(S) \dif t.
\eea
\end{subequations}
The term $ {\cal F}_\varTheta(\varTheta)$ corresponds to diabatic terms such as the molecular diffusion process or the radiative heat transfer.

\subsection{Conservation of momentum}

To derive a stochastic representation of the Navier-Stokes equations, pressure forcing is decomposed into continuous component, $p$, and white-noise term $\dot{p}_{\sigma} = {\dif_t p_{\sigma}}/{\dif t}$. The smooth component of the velocity is not only assumed continuous but also time-differentiable \citep{Memin14}. 
As demonstrated in Appendix \ref{Appendix Stochastic Navier-Stokes model}, the flow dynamics for an observer in an uniformly rotating coordinate frame writes:
\\\!\\
\fcolorbox{black}{lightgray}{
\begin{minipage}{0.95\textwidth}
\begin{center}
\bf 
Navier-Stokes equations under location uncertainty in a rotating frame
\end{center}
\begin{subequations}
\label{sto-NS-Rot}
\begin{align}
&\!\!\text{\em Momentum equations} \nonumber\\
&\;\;\;\;
\label{Navier Stokes:momentum:variation finie}
\partial_t \w 
+ \left( \w^* \bcdot \nab \right) \w
-  \frac{1}{2\rho} \sum_{i,j}  \partial_{i}\biggl (\rho a_{ij} \partial_j \w \biggr) 
+ \mbs f \times \w =  \mbs g - \frac{1}{\rho}\nab p  + \frac{1}{\rho}{\cal F}(\w),\\
&\!\!\text{\em Effective drift} \nonumber\\
&\;\;\;\;\w^*= \w - \tfrac{1}{2} (\nab\bcdot \mbs a)\transp,
\\
&\!\!\text{\em  Random pressure contribution} \nonumber\\
&\;\;\;\;
\label{Navier Stokes:momentum:martingale}
\nab \dif_t p_{\sigma}  =\!
 \left( \bsigma \dif \B_t  \bcdot \nab \right) \w
 - \rho \mbs f \times \bsigma \dif \B_t + {\cal F}(\bsigma \dif\B_t),\\
&\!\!\text{\em Mass conservation} \nonumber\\
&\;\;\;\;\Dt_t \rho + \rho \nab\bcdot \w^* \dif t=0, \;\;{\nab} \bcdot(\bsigma \dif\B_t ) =0.
\end{align}
\end{subequations}
\end{minipage}
} 
\\\!\\\!\\
Similarly to the Reynolds decomposition, the dynamics associated with the  drift component includes an additional stress term, 
and the large-scale velocity component is advected by an eddy effective drift velocity. 
The density is driven by a stochastic mass conservation equation or alternatively through the stochastic transport of temperature and salinity (\ref{transportTemp}-\ref{transportSalinity}), together with a state law. The random density constitutes a random forcing in the large-scale momentum equation. 

For incompressible flows, 
the pressure is then recovered from a modified Poisson equation; 
\begin{equation}
\label{Sto-Pressure-Poisson}
- \Delta p= \nab\bcdot \biggl(\rho  \bigl(\w^*\bcdot \nab \bigr) \w + \rho \mbs f \times \w  - \tfrac{1}{2}\sum_{ij} \partial_i (\; \rho a_{ij}  \partial_j\w)\biggr).
\end{equation}
The pressure acts as a Lagrangian penalty term  to constrain the large scale component to be divergent-free. 

This formalization can be compared to another stochastic framework based on scale gap: Stochastic Super-Parametrization (SSP) \citep{Grooms13,Grooms14}. Both modeling enable separating the large-scale velocity \eqref{Navier Stokes:momentum:variation finie} and the small-scale contribution \eqref{Navier Stokes:momentum:martingale}. This is done by a differentiability assumption on the large-scale drift, $\w$, in the modeling under location uncertainty, and through the Reynolds decomposition and a point approximation assumption in SSP. However, it can be pointed out that  no averaging procedure is settled in the modeling under location uncertainty. Furthermore, the transports of density, temperature and salinity involve random forcings. Unlike SSP, the whole system to be simulated is thus random. This randomness is of main importance for Uncertainty Quantification (UQ) aplications as illustrated theoretically in section \ref{Transport under location uncertainty} and numerically in the part II of this set of papers \citep{resseguier2016geo2}. Another main difference between the two methods lies in the subgrid tensors parametrization. Each SSP scalar evolution law involves a different subgrid tensor whereas there is a single one (related to the small-scale velocity) for every transports under location uncertainty. For both model it can be noted that the small-scale velocity component is Gaussian conditionally on the large-scale properties. Unlike our models, the SSP proposes a simple evolution model for this unresolved velocity and hence for its statistics. This type of linear forced-dissipative evolution laws, introduced by Eddy-Damped Quasi Normal Markovian (EDQNM) models \citep{orszag1970analytical,Leith71,chasnov1991simulation}, could be  as well used  to specify the diffusion operator $\bsigma$ and close the models under location uncertainty. Yet, such closure also need to be parametrized.

\subsection{Atmosphere and Ocean dynamics approximations}

Ocean and atmosphere dynamical models generally rely on several successive approximations. In the following, we review these approximations within the uncertainty framework.  

For ocean and atmosphere flows, a partition of the density and pressure is generally considered:
\begin{subequations}
\label{p-rho-decomposition}
\bea
\rho &=& \rho_b + \rho_0(z) + \rho'(x,y,z,t),\\
p &=&  \widetilde p(z) + p'(x,y,z,t).
\eea
\end{subequations}
Fields $\widetilde \rho(z)= \rho_b  + \rho_0(z)$ and $\widetilde p(z)$ correspond to the density and the pressure at equilibrium (without any motion), respectively; they are deterministic functions and depend on the height only. The pressure and density departures, $p'$ and $\rho'$, are random functions, depending on the uncertainty component. From the expression of the vertical velocity component (\ref{sto-NS-Rot}a), the equilibrium fields are related through an hydrostatic balance:
\begin{equation}
\label{H-bal}
\frac{\partial \widetilde p}{\partial z} = - g\widetilde \rho(z).
\end{equation}

\subsubsection{Traditional approximation}
This approximation helps to neglect the deflecting rotation forces associated with vertical movements. Considering the first moment conservation  along the vertical direction of (\ref{sto-NS-Rot}), with the hydrostatic balance (\ref{H-bal}), it writes: 
\begin{equation}
\partial_t w 
+  \left( \w^* \bcdot \nab \right) w 
-  \tfrac{1}{2} \sum_{i,j}  \partial_{i}\biggl (a_{ij} \partial_j w\biggr)  
+ f_x v - f_y u 
=  
- \frac{1}{\rho}\left [\rho' g+ \frac{\partial p'}{\partial z}\right] 
+ {\cal F}(w).
\end{equation}
This approximation is justified when an hydrostatic assumption is employed.

\subsubsection{Boussinesq approximation}

Within small density fluctuations ({\rm i.e.} the Boussinesq approximation) as observed in the ocean, the stochastic mass conservation reads  
\begin{equation}
\label{approxBoussinesq_isochoric}
0 = \Dt_t \rho + \rho \nab\bcdot \w^*  \dif t
 \approx \rho_b \nab\bcdot \w^*  \dif t.
 \end{equation}
This implies that the flow is volume-preserving. In an anelastic approximation, density variations dominate. It can be shown we get the weaker constraint, associated with an horizontal uncertainty:
\bea 
\nab\bcdot \w - \tfrac{1}{2} \nab_H\bcdot(\nab_H\bcdot \mbs{a}_H)\transp = \frac{g}{c^2\rho}(w\tilde \rho)
\eea 
where $c^{-2}$ denotes the velocity of the acoustic waves and subscript $H$ indicates the set of horizontal coordinates. The classical anelastic constraint implicitly assumes a divergence-free condition on the variance tensor divergence (as obtained for homogeneous turbulence). 

According to equations \eqref{transportTemp} and \eqref{transportSalinity}, temperature and salinity are transported by the random flow. If those tracers do not oscillate too much, the density anomaly, $\rho - \rho_b$, can be approximated by a linear combination of these two properties. And thus, in the Boussinesq approximation, this anomaly is transported:
\begin{equation}
\label{Rhoz}
0 = \rm D_t (\rho-\rho_b) = \Dt_t (\rho-\rho_b) .
 \end{equation}
\cite{Holm2015} obtained the very same stochastic transport of density anomaly from a Lagrangian mechanics approach.

Using the same approximation, the contribution of the momentum material derivative associated with the density variation can be neglected.
The Navier-Stokes equations coupling the Boussinesq and traditional approximations then read: 
\\\!\\
\fcolorbox{black}{lightgray}{
\begin{minipage}{0.95\textwidth}
\begin{center}
\bf  Simple Boussinesq equations under location uncertainty
\end{center}
\begin{subequations}
\label{sto-Boussinesq-buo1}
\begin{align}
&\!\!\text{\em Momentum equations} \nonumber\\
&\;\;\;\; 
\partial_t \w
+ \left( \w^* \bcdot \nab \right) \w
-  \tfrac{1}{2} \sum_{i,j}  \partial_{i}\biggl (a_{ij} \partial_j \w \biggr) 
+  f \mbs k  \times \mbs u
=
b \; \mbs k
- \frac{1}{\rho_b}\nab p'  
+ {\cal F}(\w),
%
\label{w-comp}\\
&\!\!\text{\em Effective drift} \nonumber\\
&\;\;\;\;
\w^*
=
\begin{pmatrix}
\mbs u^* \\ w^*
\end{pmatrix}
=
\w - 
\tfrac{1}{2} (\nab\bcdot \mbs a)\transp ,\\
&\!\!\text{\em Buoyancy equation} \nonumber\\
&\;\;\;\;
\Dt_t b  
+ N^2 \left( w^* \dif t  
+ (\bsigma \dif \B_t)_z \right)=
\tfrac{1}{2} \nab \bcdot \left( \mbs a_{\bullet z} N^2 \right) \dif t,
 \label{eq-buo}\\
&\!\!\text{\em Random pressure fluctuation} \nonumber\\
&\;\;\;\;
\nab \dif_t p_{\sigma}  =\!
 -  \rho_b \left ( \bsigma \dif \B_t \bcdot \nab \right ) \w^*
 - f\mbs k  \times (\bsigma \dif \B_t)_H+ {\cal F}(\bsigma \dif \B_t),\label{dp}\\
&\!\!\text{\em Incompressibility} \nonumber\\
&\;\;\;\;  \nab \bcdot \w ={\nab} {\bcdot}\bigl(\bsigma \dot{ \B}\bigr)  = \nab\bcdot\nab \bcdot \mbs a =0.
\label{inc-cond-Boussinesq}
\end{align}
\end{subequations}
\end{minipage}
} 
\\\!\\\!\\
For this system, the thermodynamics equations are expressed through the buoyancy variable $b=-g \rho'/\rho_b$, and the stratification (Brunt-V\"ais\"al\"a frequency) $N^2(z)=-g /{\rho_b}\  \partial_z \rho_0(z)$ is introduced. The buoyancy term constitutes a random forcing of the vertical large-scale velocity  component.
Since the density anomaly, $\rho-\rho_b$, has been decomposed into a constant background slope and a residual, the multiplicative noise of equation \eqref{Rhoz} is split into an additive and a multiplicative noise in \eqref{eq-buo}. The additive noise drains random energy from the stratification toward the buoyancy. Therefore, the buoyancy energy is not conserved due to the background stratification.

\subsubsection{Buoyancy oscillations}


 To illustrate the effect of this additive noise in simple cases, we consider here constant-along-depth buoyancy anomaly and stratification  ($\partial_z b = 0$ and $\partial_z N =0$) and only a vertical motion component  ({\em i.e} $\mbs u =0$ and $(\bsigma\dif\B_t)_H=0$) with no dependence on depth (due to the divergence constraint). Note that this latter constraint on the diffusion tensor, implies that only $a_{zz}$ is non null with no dependence on depth as well. Then, the Boussinesq equations read
 \bea
\partial_t w = b  \text{ and }
\dif_t b = - N^2 (w \dif t +(\bsigma\dBt)_{z}).
\label{buo-strat}
\eea 
Similarly to the deterministic case, we recognize an oscillatory system  if $N^2>0$ and a diverging system if $N^2<0$ ({\rm i.e.} when lighter fluid is below heavier fluid). The velocity and buoyancy are coupled by gravity and transport. However, in our stochastic framework, the density anomaly is also transported by a random velocity. 
This highly oscillating velocity may be interpreted as the action of wind on the surface of the ocean.
The interaction between this unresolved velocity component and the stratification acts has a random forcing on the oscillator:
\begin{equation}
\dif_t \partial_t w  + N^2 w \dif t =-N^2 (\bsigma\dBt)_{z}.
\label{Oscillator randomly forced}
\end{equation}
To solve this equation, one can note that:
\bea
\dif_t \left( {\rm e}^{-2\rm i Nt} \partial_t ( {\rm e}^{\rm i Nt} w ) \right) = - N^2 {\rm e}^{-\rm i Nt}  (\bsigma \dBt )_z .
\eea
Then, by integrating twice, we get the solutions of the stochastic system \eqref{buo-strat}:
 \begin{align}
w(t)  &= 
\underbrace{
w(0) \cos(Nt) +  \partial_t w(0)/N \sin(Nt) 
}_{=\Exp (w(t) )}
- N\int_{0}^t \sin\bigl(N(t-r)\bigr) (\bsigma\dif \B_r)_{z},\\
b(t)  &=  
\underbrace{
\partial_t w(0)  \cos(Nt) -w(0) N \sin(Nt)
}_{=\Exp (b(t) )}
-N^2 \int_0^t  \cos(N(t-r)) (\bsigma\dif \B_r)_{z}.
\end{align}
The ensemble means are the traditional deterministic solutions whereas the random parts are continuous summations of sine wave with uncorrelated random amplitudes.
At each time $r$, the additive random forcing introduces an oscillation. Without dissipative processes, the latter remains in the system. But, the influence of the past excitations are weighed by sine wave due to the phase change. The buoyancy and the velocity are Gaussian random variables (as linear combinations of independent Gaussian variables). Therefore, their finite dimensional law ({\em i.e}. the  multi-time probability density function) are entirely defined by their mean and covariance functions.
 The variances can be computed through the Ito isometry \citep{Oksendal98}. Then, the velocity covariance can be inferred from the SDE \eqref{Oscillator randomly forced}:
\bea
 \label{cov_buo-strat}
Cov_w(t, t+\tau) 
 =
  \frac{a_{zz}N} 4 
 \cos(N \tau ) \left ( 2 N t - \sin( 2 N t) \right )
 + \frac{ a_{zz}N} 4 \sin(N \tau ) \left ( 1 - \cos( 2 N t) \right )
 .
 \eea
The covariance of the buoyancy is similar. Since the interaction between the unresolved velocity component and the background density gradient cannot be resolved deterministically, uncertainties of the dynamics accumulate. Each time introduces a new random uncorrelated excitation. This is why the buoyancy and velocity variances increase linearly with time.
In contrast, in a deterministic oscillator with random perturbations of the initial conditions, the variance remains constant and depends solely on the initial velocity variance.
This growing also illustrates in a very simple case the possible destabilization effects of the unresolved velocity in the models under location uncertainty.

The first term of the covariance \eqref{cov_buo-strat} modulates the variance with a sine wave. The randomness of $w$ is generated by a set of sine wave which have coherent phases and interfere. When $N \tau = 0 [2 \pi]$ the noises with correlated amplitudes, $(\bsigma\dif \B_r)_{z}$, in $w(t)$ and $w(t+\tau)$ are in phase, and thus the velocity covariance is large. When $N \tau = \pi [2 \pi]$ these correlated noises have opposite phases, and yields a negative velocity covariance. When $N \tau$ is close to $ \frac \pi 2 [ \pi]$, the noises are in quadrature and the first term of the velocity covariance is zero.

\subsection{Summary}

The fundamental conservation laws (mass, momentum and energy) have been interpreted within the proposed stochastic framework. Usual approximations of fluid dynamics are considered, leading to a stochastic version of Boussinesq equations. As developed, the buoyancy is transported by a smooth large-scale velocity component and a small-scale random field, delta-correlated in time. Consequently, the buoyancy is forced by an additive and a multiplicative noises, uncorrelated in time but correlated in space. The additive noise encodes the interaction between the unresolved velocity and the background stratification. The resulting random buoyancy then appears as an additive time-correlated random forcing in the vertical momentum equation. 
Both momentum and thermodynamic equations then involve an inhomogeneous and anisotropic diffusion, and a drift correction that both depend on the unresolved velocity variance tensor, $\mbs a$. Assuming hydrostatic equilibrium in this stochastic Boussinesq model directly provides a stochastic version of the primitive equations. A solvable model is also derived from this Boussinesq model. This toy model exemplifies how the random forcing continually increases the variance of the solution.

\subsection{Guidelines for the derivation of models under location uncertainty}
\label{Guidelines for the derivation of models under location uncertainty}
The main steps of the derivation of dynamics under location uncertainty are sketched out below. 
\begin{enumerate}[(i)]
\item The conservation laws of classical mechanics describe variation of some extensive properties. As illustrated in Appendix \ref{Appendix Stochastic Navier-Stokes model} for the stochastic Navier-Stokes model, if the extensive property of interest (linear momentum in this Appendix) has a component uncorrelated in time, the variations of this component must be balanced by a very irregular forcing, and can be discarded.
\item The stochastic Reynolds transport theorem \eqref{th_transport} enables us to interpret the variation of the time-correlated component of the extensive property. The expression of the stochastic material derivative of an associated intensive quantity follows.
\item The formulas \eqref{link DD and material deriv} relate this material derivative, $\rm D_t$, to the stochastic transport operator, $\Dt_t$. In most cases, these operators coincide.
\item Gathering the equations from (ii) and (iii) provides an explicit Eulerian evolution law.
\item Additional regularity assumptions can be used to separate the large-scale and small-scale components of the evolution law. As an example, the velocity component, $\w$, has been assumed to be differentiable with respect to time in this section {\rm i.e.} the acceleration component, $\partial_t \w$, is correlated in time. Thus, there is no time-uncorrelated noise in the large-scale momentum evolution law and the random pressure fluctuations appear in a separate equation. This separation is of great interest for deterministic LES-like simulations. However, by this approximation, we lose the conservation of the kinetic energy \eqref{E-cons}. For Uncertainty Quantification (UQ) purposes, this separation is not necessary.
\item With or without regularity assumptions, usual approximations ({\em e.g.} the Boussinesq approximation) can be done to simplify further the stochastic model.
\end{enumerate}
Let us point out that the corresponding models involve subgrid terms which generally cannot be neglected.
When adimentionalized,  those subgrid terms are weighted by an additional adimentional number whose value depends on the noise magnitude. For a low noise the approximate dynamical models  take a random form that remains similar  to their deterministic counterparts. At the opposite, the system is generally significantly changed when considering  a strong noise.

A second companion paper (part II) \citep{resseguier2016geo2} describes random versions of Quasi-Geostrophic (QG) and Surface Quasi-Geostrophic (SQG) models with a moderate influence of the subgrid terms, whereas the third one (part III) \citep{resseguier2016geo3} focuses on the same models with a stronger influence of subgrid terms. The two dynamics are significantly different.
 \\


To close the stochastic system, the operator $\bsigma$ needs to be fully specified. Several solutions can be proposed to that purpose. The simplest specification consists in resorting to a homogeneous parametrization such as the Kraichnan model \citep{Kraichnan68,kraichnan1994anomalous,Gawedzki95,Majda-Kramer}. The companion paper \cite{resseguier2016geo2} relies on this type of random field with a parameterization fixed from an ideal spectrum. When the small-scale velocity is observable or at least partially observable the structure of that operator can then be estimated. For instance, in \cite{resseguier2015reduced} a nonparametric and inhomogeneous variance tensor $\mbs a (\xx) = \bsigma (\xx) \bsigma (\xx)\transp$ is estimated from a sequence of observed velocity. Parametric and/or homogeneous models could also be specified. If no small-scale statistics are available, the choice of a closure can expressed $\bsigma$ as a function of large-scale quantities and similarity assumption \citep{Kadri-Memin-16,Chandramouli16}. The unresolved velocity can be defined as the solution of a simple linearized equations subject to advection by large-scale components, damping and additive random forcing as in {\em e.g.} quasi-linear approximations \citep{Farrell14} or stochastic super-parameterizations \citep{Grooms13,Grooms14}. Existing methodologies of data assimilation literature would also be of great interest in this context. Several authors define models from observed correlation length or correlation deformation estimation \citep{pannekoucke2008estimation,mirouze2010representation,Weaver-Courtier01}. Others specify the correlation matrices by diffusion equations \citep{michel2013estimatinga,michel2013estimatingb,pannekoucke2014modelling}.

\section{Conclusion}

In this paper, a random component is added to the smooth velocity field. This helps model a coarse-grained effect. The random component 
is chosen Gaussian and uncorrelated in time. Nevertheless, it can be inhomogeneous and anisotropic in space. 
With such a velocity, the expression of the material derivative is changed. To explicit this change, we introduce the stochastic transport operator, $\Dt_t$. The material derivative, $\rm D_t$, generally coincides with this operator, especially for tracer transports. Otherwise, the difference between these operators has a simple analytic expression. The stochastic transport operator involves an anisotropic and inhomogeneous diffusion, a drift correction and a multiplicative noise. These terms are specified by the statistics of the sub-grid velocity. The diffusion term generalizes the Boussinesq assumption. Moreover, the link between the three previous terms ensures many desired properties for tracers, such as energy conservation and continuous variance increasing. For passive tracer, the PDEs of mean and variance field are derived. The unresolved velocity transfers energy from the small-scale mean field to the variance. This is very suitable to quantify the uncertainty associated with sub-grid dynamics. This randomized dynamics has been called transport under location uncertainty. A stochastic version of the Reynolds transport theorem is then derived. It enables us to compute the time differentiation of extensive properties to interpret the conservation laws of classical mechanics in a stochastic sense.

Applied to the conservation of linear momentum, amount of substance and first principle of thermodynamics, a stochastic version of the Navier-Stokes equations is obtained. Similarly to the deterministic case, a small buoyancy assumption leads to random Boussinesq equations. The random transport of buoyancy involves both a multiplicative and an additive noises. The additive noise encodes the interaction between the unresolved velocity and the background stratification. We schematically presented the action of this last forcing through a solvable model of fluid parcels vertical oscillations.
\\


Under strong rotation and strong stratification assumptions, the stochastic Boussinesq representation simplifies to different mesoscale models depending on the scaling of the subgrid terms. The companion papers part II \citep{resseguier2016geo2} and part III \citep{resseguier2016geo3} describe such models. For a moderate influence of noise-driven subgrid terms, the Potential Vorticity (PV) is randomly transported up to three source terms \citep{resseguier2016geo2}. Assuming zero PV in the fluid interior yields the usual Surface Quasi-Geostrophic (SQG) relationship. The stochastic transport of buoyancy, yields a stochastic SQG model referred to as SQG model under Moderate Uncertainty ($SQG_{MU}$). This two-dimensional nonlinear dynamics enables \cite{resseguier2016geo2} to numerically unveil advantages of the models under location uncertainty in terms of small-scale structures restoration (in a single realization) and ensemble model error prediction (with an improvement compared to perturbed deterministic models of one order of magnitude).

To go beyond the framework of this paper, larger-scale random dynamics can be inferred by averaging the models under location uncertainty using singular perturbation or stochastic invariant manifold theories \citep{gottwald2013role}. Finally, a delta-correlated process and stochastic calculus may seem insufficient to model the smallest velocity scales. Ito formulas deal with white-noise forcing and contains only second-order terms. For higher order terms, such as hyperviscosity, more complete theories exist \citep{klyatskin2005stochastic}.


\section*{Acknowledgments}


The authors thank Guillaume Lapeyre, Aur\'elien Ponte, Jeroen Molemaker, Guillaume Roulet and Jonathan Gula for helpful discussions. We also acknowledge the support of the ESA DUE GlobCurrent project (contract no. 4000109513/13/I-LG), the ``Laboratoires d'Excellence''  CominLabs, Lebesgue and Mer (grant ANR-10-LABX-19-01) through the SEACS project.


\bibliographystyle{plainnat}

\bibliography{biblio}

\newcommand{\noop}[1]{}
\begin{thebibliography}{66}
\providecommand{\natexlab}[1]{#1}
\providecommand{\url}[1]{\texttt{#1}}
\expandafter\ifx\csname urlstyle\endcsname\relax
  \providecommand{\doi}[1]{doi: #1}\else
  \providecommand{\doi}{doi: \begingroup \urlstyle{rm}\Url}\fi

\bibitem[Allen and Stainforth(2002)]{allen2002towards}
Myles Allen and David Stainforth.
\newblock Towards objective probabilistic climate forecasting.
\newblock \emph{Nature}, 419\penalty0 (6903):\penalty0 228--228, 2002.

\bibitem[Anderson and Anderson(1999)]{Anderson99}
J.L. Anderson and S.L. Anderson.
\newblock A {M}onte {C}arlo implementation of the nonlinear filtering problem
  to produce ensemble assimilations and forecasts.
\newblock \emph{Mon. Weather Rev.}, 127\penalty0 (12):\penalty0 2741--2758,
  1999.

\bibitem[Berner et~al.(2011)Berner, Ha, Hacker, Fournier, and
  Snyder]{berner2011model}
J~Berner, S-Y Ha, J~Hacker, A~Fournier, and C~Snyder.
\newblock Model uncertainty in a mesoscale ensemble prediction system:
  Stochastic versus multiphysics representations.
\newblock \emph{Mon. Weather Rev.}, 139\penalty0 (6):\penalty0 1972--1995,
  2011.

\bibitem[Berner et~al.(2015)Berner, Achatz, Batte, Camara, Crommelin,
  Christensen, Colangeli, Dolaptchiev, Franzke, Friederichs, Imkeller,
  Jarvinen, Juricke, Kitsios, Lott, Lucarini, Mahajan, Palmer, Penland, Storch,
  Sakradzija, Weniger, Weisheimer, Williams, and Yano]{Berner15}
J.~Berner, U.~Achatz, L.~Batte, A.~De~La Camara, D.~Crommelin, H.~Christensen,
  M.~Colangeli, S.~Dolaptchiev, C.L.E. Franzke, P.~Friederichs, P.~Imkeller,
  H.~Jarvinen, S.~Juricke, V.~Kitsios, F.~Lott, V.~Lucarini, S.~Mahajan, T.~N.
  Palmer, C.~Penland, J.-S.~Von Storch, M.~Sakradzija, M.~Weniger,
  A.~Weisheimer, P.~D. Williams, and J.-I. Yano.
\newblock Stochastic parameterization: Towards a new view of weather and
  climate models.
\newblock Technical report, arXiv:1510.08682 [physics.ao-ph], 2015.

\bibitem[Bocquet et~al.(2016)Bocquet, Gurumoorthy, Apte, Carrassi, Grudzien,
  and Jones]{bocquet2016degenerate}
M.~Bocquet, K.~Gurumoorthy, A.~Apte, A.~Carrassi, C.~Grudzien, and C.~Jones.
\newblock Degenerate kalman filter error covariances and their convergence onto
  the unstable subspace.
\newblock \emph{arXiv preprint arXiv:1604.02578}, 2016.

\bibitem[Brankart(2013)]{Brankart13}
J.~Brankart.
\newblock Impact of uncertainties in the horizontal density gradient upon low
  resolution global ocean modeling.
\newblock \emph{Ocean Model.}, 66:\penalty0 64--76, 2013.

\bibitem[Buizza et~al.(1999)Buizza, Miller, and Palmer]{Buizza99}
R.~Buizza, M.~Miller, and T.N. Palmer.
\newblock Stochastic representation of model uncertainties in the ecmwf
  ensemble prediction system.
\newblock \emph{Q. J. Roy. Meteor. Soc.}, 125:\penalty0 2887--2908, 1999.

\bibitem[Chandramouli et~al.(2016)Chandramouli, Heitz, Laizet, and
  M{\'e}min]{Chandramouli16}
P.~Chandramouli, D.~Heitz, S.~Laizet, and E.~M{\'e}min.
\newblock Coarse-grid large eddy simulations in a wake flow with new stochastic
  small scale models.
\newblock ArXiv report, 2016.

\bibitem[Chapron et~al.(2005)Chapron, Collard, and Ardhuin]{chapron2005direct}
Bertrand Chapron, Fabrice Collard, and Fabrice Ardhuin.
\newblock Direct measurements of ocean surface velocity from space:
  Interpretation and validation.
\newblock \emph{J. Geophys. Res.-Oceans}, 110\penalty0 (C7), 2005.
\newblock ISSN 2156-2202.
\newblock C07008.

\bibitem[Chasnov(1991)]{chasnov1991simulation}
Jeffrey~R Chasnov.
\newblock Simulation of the kolmogorov inertial subrange using an improved
  subgrid model.
\newblock \emph{Phys. Fluids A}, 3\penalty0 (1):\penalty0 188--200, 1991.

\bibitem[{Da Prato} and Zabczyk(1992)]{DaPrato}
G.~{Da Prato} and J.~Zabczyk.
\newblock \emph{Stochastic equations in infinite dimensions}.
\newblock Cambridge University Press, 1992.

\bibitem[Farrell and Ioannou(2014)]{Farrell14}
B.~Farrell and P.~Ioannou.
\newblock Statistical state dynamics: a new perspective on turbulence in shear
  flow.
\newblock \emph{arXiv preprint arXiv:1412.8290}, 2014.

\bibitem[Flandoli(2011)]{flandoli2011interaction}
F.~Flandoli.
\newblock The interaction between noise and transport mechanisms in pdes.
\newblock \emph{Milan J. Math.}, 79\penalty0 (2):\penalty0 543--560, 2011.

\bibitem[Franzke and Majda(2006)]{Franzke06}
C.~Franzke and A.~Majda.
\newblock Low-order stochastic mode reduction for a prototype atmospheric gc.
\newblock \emph{J. Atmos. Sci.}, pages 457--479, 2006.

\bibitem[Franzke et~al.(2015)Franzke, O'Kane, Berner, Williams, and
  Lucarini]{Franzke15}
C.~Franzke, T.~O'Kane, J.~Berner, P.~Williams, and V.~Lucarini.
\newblock Stochastic climate theory and modeling.
\newblock \emph{WIREs Clim. Change}, 6\penalty0 (1):\penalty0 63--78, 2015.

\bibitem[Franzke et~al.(2005)Franzke, Majda, and Vanden-Eijnden]{Franzke05}
Christian Franzke, Andrew~J Majda, and Eric Vanden-Eijnden.
\newblock Low-order stochastic mode reduction for a realistic barotropic model
  climate.
\newblock \emph{J. Atmos. Sci.}, 62\penalty0 (6):\penalty0 1722--1745, 2005.

\bibitem[Gawedzky and Kupiainen(1995)]{Gawedzki95}
K.~Gawedzky and A.~Kupiainen.
\newblock Anomalous scaling of the passive scalar.
\newblock \emph{Phys. Rev. Lett.}, 75:\penalty0 3834--3837, 1995.

\bibitem[Gent and McWilliams(1990)]{Gent90}
P.~Gent and J.~McWilliams.
\newblock Isopycnal mixing in ocean circulation models.
\newblock \emph{J. Phys. Oceanogr.}, 20:\penalty0 150--155, 1990.

\bibitem[Gottwald and Harlim(2013)]{gottwald2013role}
G.~Gottwald and J.~Harlim.
\newblock The role of additive and multiplicative noise in filtering complex
  dynamical systems.
\newblock \emph{Proc. R. Soc. A}, 469\penalty0 (2155):\penalty0 20130096, 2013.
\newblock ISSN 1364-5021.

\bibitem[Gottwald and Melbourne(2013)]{Gottwald13}
G.~Gottwald and I.~Melbourne.
\newblock Homogenization for deterministic maps and multiplicative noise.
\newblock \emph{Proc. R. Soc. A}, 469:\penalty0 20130201, 2013.

\bibitem[Grooms and Majda(2013)]{Grooms13}
I.~Grooms and A.~Majda.
\newblock Efficient stochastic superparameterization for geophysical
  turbulence.
\newblock \emph{PNAS}, 110\penalty0 (12):\penalty0 4464--4469, 2013.

\bibitem[Grooms and Majda(2014)]{Grooms14}
I.~Grooms and A.~Majda.
\newblock Stochastic superparameterization in quasigeostrophic turbulence.
\newblock \emph{J. Comput. Phys.}, 271:\penalty0 78--98, 2014.

\bibitem[Hasselmann(1976)]{Hasselmann76}
K.~Hasselmann.
\newblock Stochastic climate models. part i: theory.
\newblock \emph{Tellus}, 28:\penalty0 473--485, 1976.

\bibitem[Holm(2015)]{Holm2015}
Darryl~D. Holm.
\newblock Variational principles for stochastic fluid dynamics.
\newblock \emph{Proc. R. Soc. A}, 471\penalty0 (2176):\penalty0 20140963, 2015.
\newblock ISSN 1364-5021.
\newblock \doi{10.1098/rspa.2014.0963}.

\bibitem[Kadri-Harouna and M{\'e}min(2016)]{Kadri-Memin-16}
S.~Kadri-Harouna and E.~M{\'e}min.
\newblock Stochastic representation of the {R}eynolds transport theorem:
  revisiting large-scale modeling.
\newblock Manuscript submitted for publication, 2016.

\bibitem[Keating et~al.(2011)Keating, Smith, and Kramer]{keating2011diagnosing}
Shane Keating, Shafer Smith, and Peter Kramer.
\newblock Diagnosing lateral mixing in the upper ocean with virtual tracers:
  Spatial and temporal resolution dependence.
\newblock \emph{J. Phys. Oceanogr.}, 41\penalty0 (8):\penalty0 1512--1534,
  2011.

\bibitem[Klyatskin(2005)]{klyatskin2005stochastic}
V.~Klyatskin.
\newblock \emph{Stochastic equations through the eye of the physicist: Basic
  concepts, exact results and asymptotic approximations}.
\newblock Elsevier, 2005.

\bibitem[Kraichnan(1968)]{Kraichnan68}
R.~Kraichnan.
\newblock Small-scale structure of a scalar field convected by turbulence.
\newblock \emph{Phys. of Fluids}, 11:\penalty0 945--963, 1968.

\bibitem[Kraichnan(1994)]{kraichnan1994anomalous}
Robert Kraichnan.
\newblock Anomalous scaling of a randomly advected passive scalar.
\newblock \emph{Phys. Rev. Lett.}, 72\penalty0 (7):\penalty0 1016, 1994.

\bibitem[Kunita(1990)]{Kunita}
H.~Kunita.
\newblock \emph{Stochastic flows and stochastic differential equations}.
\newblock Cambridge University Press, 1990.

\bibitem[Leith(1971)]{Leith71}
C.~Leith.
\newblock Atmospheric predictability and two-dimensional turbulence.
\newblock \emph{J. Atmos. Sci}, 28:\penalty0 145--161, 1971.

\bibitem[Majda and Kramer(1999)]{Majda-Kramer}
A.~Majda and P.~Kramer.
\newblock Simplified models for turbulent diffusion: Theory, numerical
  modelling, and physical phenomena.
\newblock \emph{Phys. Rep.}, 314:\penalty0 237--574, 1999.

\bibitem[Majda et~al.(1999)Majda, Timofeyev, and Vanden-Eijnden]{Majda99}
A.~Majda, I.~Timofeyev, and E.~Vanden-Eijnden.
\newblock Models for stochastic climate prediction.
\newblock \emph{PNAS}, 96\penalty0 (26):\penalty0 14687--14691, 1999.

\bibitem[Majda et~al.(2001)Majda, Timofeyev, and Vanden-Eijnden]{Majda01}
A.~Majda, I.~Timofeyev, and E.~Vanden-Eijnden.
\newblock A mathematical framework for stochastic climate models.
\newblock \emph{Commun. on Pure and Applied Math.}, 54:\penalty0 891--974,
  2001.

\bibitem[Majda et~al.(2003)Majda, Timofeyev, and Vanden-Eijnden]{Majda03}
A.~Majda, I.~Timofeyev, and E.~Vanden-Eijnden.
\newblock A systematic strategies for stochastic mode reduction in climate.
\newblock \emph{Journ. Atmos. Sci.}, 60:\penalty0 1705--1722, 2003.

\bibitem[Majda(2015)]{majda2015statistical}
Andrew~J Majda.
\newblock Statistical energy conservation principle for inhomogeneous turbulent
  dynamical systems.
\newblock \emph{PNAS}, 112\penalty0 (29):\penalty0 8937--8941, 2015.

\bibitem[Mana and Zanna(2014)]{mana2014toward}
PierGianLuca Mana and Laure Zanna.
\newblock Toward a stochastic parameterization of ocean mesoscale eddies.
\newblock \emph{Ocean Model.}, 79:\penalty0 1--20, 2014.

\bibitem[Mason and Thomson(1992)]{Mason92}
P.J. Mason and D.J. Thomson.
\newblock Stochastic backscatter in large-eddy simulations of boundary layers.
\newblock \emph{J. of Fluid Mech.}, 242:\penalty0 51--78, 1992.

\bibitem[Melbourne and Stuart(2011)]{Melbourne11}
I.~Melbourne and A.M. Stuart.
\newblock A note on diffusion limits of chaotic skew-product flows.
\newblock \emph{Nonlinearity}, 24:\penalty0 1361--1367, 2011.

\bibitem[M{\'e}min(2014)]{Memin14}
E.~M{\'e}min.
\newblock Fluid flow dynamics under location uncertainty.
\newblock \emph{Geophys. Astro. Fluid}, 108\penalty0 (2):\penalty0 119--146,
  2014.

\bibitem[Michel(2013{\natexlab{a}})]{michel2013estimatingb}
Y~Michel.
\newblock Estimating deformations of random processes for correlation
  modelling: methodology and the one-dimensional case.
\newblock \emph{Q. J. Roy. Meteor. Soc.}, 139\penalty0 (672):\penalty0
  771--783, 2013{\natexlab{a}}.

\bibitem[Michel(2013{\natexlab{b}})]{michel2013estimatinga}
Yann Michel.
\newblock Estimating deformations of random processes for correlation modelling
  in a limited area model.
\newblock \emph{Q. J. Roy. Meteor. Soc.}, 139\penalty0 (671):\penalty0
  534--547, 2013{\natexlab{b}}.

\bibitem[Mikulevicius and Rozovskii(2004)]{mikulevicius2004stochastic}
R.~Mikulevicius and B.~Rozovskii.
\newblock Stochastic {N}avier--{S}tokes equations for turbulent flows.
\newblock \emph{SIAM J. Math. Anal.}, 35\penalty0 (5):\penalty0 1250--1310,
  2004.

\bibitem[Mirouze and Weaver(2010)]{mirouze2010representation}
I~Mirouze and AT~Weaver.
\newblock Representation of correlation functions in variational assimilation
  using an implicit diffusion operator.
\newblock \emph{Q. J. Roy. Meteor. Soc.}, 136\penalty0 (651):\penalty0
  1421--1443, 2010.

\bibitem[Nakamura(2001)]{Nakamura01}
N.~Nakamura.
\newblock A new look at eddy diffusivity as a mixing diagnostic.
\newblock \emph{J. Atmos. Sci.}, 58\penalty0 (24):\penalty0 3685--3701, 2001.

\bibitem[Ng et~al.(2011)Ng, McLaughlin, Entekhabi, and Ahanin]{ng2011role}
G.-H. Ng, D.~McLaughlin, D.~Entekhabi, and A.~Ahanin.
\newblock The role of model dynamics in ensemble kalman filter performance for
  chaotic systems.
\newblock \emph{Tellus A}, 63\penalty0 (5):\penalty0 958--977, 2011.

\bibitem[Oksendal(1998)]{Oksendal98}
B.~Oksendal.
\newblock \emph{Stochastic differential equations}.
\newblock Spinger-Verlag, 1998.

\bibitem[Orszag(1970)]{orszag1970analytical}
S.~Orszag.
\newblock Analytical theories of turbulence.
\newblock \emph{J. Fluid Mech.}, 41\penalty0 (02):\penalty0 363--386, 1970.

\bibitem[Pannekoucke and Massart(2008)]{pannekoucke2008estimation}
Olivier Pannekoucke and S~Massart.
\newblock Estimation of the local diffusion tensor and normalization for
  heterogeneous correlation modelling using a diffusion equation.
\newblock \emph{Q. J. Roy. Meteor. Soc.}, 134\penalty0 (635):\penalty0
  1425--1438, 2008.

\bibitem[Pannekoucke et~al.(2014)Pannekoucke, Emili, and
  Thual]{pannekoucke2014modelling}
Olivier Pannekoucke, Emanuele Emili, and Olivier Thual.
\newblock Modelling of local length-scale dynamics and isotropizing
  deformations.
\newblock \emph{Q. J. Roy. Meteor. Soc.}, 140\penalty0 (681):\penalty0
  1387--1398, 2014.

\bibitem[Pavliotis and Stuart(2008)]{Pavliotis08}
G.~Pavliotis and A.M. Stuart.
\newblock \emph{Multiscale methods: Averaging and homogeneization}.
\newblock Springer, 2008.

\bibitem[Penland(2003)]{penland2003stochastic}
C{\'e}cile Penland.
\newblock A stochastic approach to nonlinear dynamics: A review (extended
  version of the article-" noise out of chaos and why it won't go away").
\newblock \emph{Bull. Amer. Meteor. Soc.}, 84\penalty0 (7):\penalty0 925--925,
  2003.

\bibitem[Poje et~al.(2010)Poje, Haza, {\"O}zg{\"o}kmen, Magaldi, and
  Garraffo]{poje2010resolution}
Andrew Poje, Angelique Haza, Tamay {\"O}zg{\"o}kmen, Marcello Magaldi, and
  Zulema Garraffo.
\newblock Resolution dependent relative dispersion statistics in a hierarchy of
  ocean models.
\newblock \emph{Ocean Model.}, 31\penalty0 (1):\penalty0 36--50, 2010.

\bibitem[Pr{\'e}v{\^o}t and R{\"o}ckner(2007)]{Prevot07}
C.~Pr{\'e}v{\^o}t and M.~R{\"o}ckner.
\newblock \emph{A concise course on stochastic partial differential equations},
  volume 1905.
\newblock Springer, 2007.

\bibitem[Resseguier et~al.(2015)Resseguier, M{\'e}min, and
  Chapron]{resseguier2015reduced}
V.~Resseguier, E.~M{\'e}min, and B.~Chapron.
\newblock Reduced flow models from a stochastic {N}avier-{S}tokes
  representation.
\newblock \emph{ISUP}, 2015.

\bibitem[Resseguier et~al.(2017{\natexlab{a}})Resseguier, M{\'e}min, and
  Chapron]{resseguier2016geo2}
V.~Resseguier, E.~M{\'e}min, and B.~Chapron.
\newblock Geophysical flows under location uncertainty, part ii:
  Quasi-geostrophic models and efficient ensemble spreading.
\newblock Manuscript submitted for publication, 2017{\natexlab{a}}.

\bibitem[Resseguier et~al.(2017{\natexlab{b}})Resseguier, M{\'e}min, and
  Chapron]{resseguier2016geo3}
V.~Resseguier, E.~M{\'e}min, and B.~Chapron.
\newblock Geophysical flows under location uncertainty, part iii: Sqg and
  frontal dynamics under strong turbulence.
\newblock Manuscript submitted for publication, 2017{\natexlab{b}}.

\bibitem[Sapsis(2013)]{sapsis2013attractor}
T.~Sapsis.
\newblock Attractor local dimensionality, nonlinear energy transfers and
  finite-time instabilities in unstable dynamical systems with applications to
  two-dimensional fluid flows.
\newblock \emph{Proc. R. Soc. A}, 469\penalty0 (2153):\penalty0 20120550, 2013.
\newblock ISSN 1364-5021.
\newblock \doi{10.1098/rspa.2012.0550}.

\bibitem[Sapsis and Majda(2013{\natexlab{a}})]{sapsis2013blending}
T.~Sapsis and A.~Majda.
\newblock Blending modified {G}aussian closure and non-{G}aussian reduced
  subspace methods for turbulent dynamical systems.
\newblock \emph{J. Nonlinear Sci.}, 23\penalty0 (6):\penalty0 1039--1071,
  2013{\natexlab{a}}.

\bibitem[Sapsis and Majda(2013{\natexlab{b}})]{sapsis2013statistically}
T.~Sapsis and A.~Majda.
\newblock Statistically accurate low-order models for uncertainty
  quantification in turbulent dynamical systems.
\newblock \emph{PNAS}, 110\penalty0 (34):\penalty0 13705--13710,
  2013{\natexlab{b}}.

\bibitem[Shutts(2005)]{Shutts05}
G.~Shutts.
\newblock A kinetic energy backscatter algorithm for use in ensemble prediction
  systems.
\newblock \emph{Q. J. Roy. Meteor. Soc.}, 612:\penalty0 3079--3012, 2005.

\bibitem[Snyder et~al.(2015)Snyder, Bengtsson, and Morzfeld]{Snyder15}
C.~Snyder, T.~Bengtsson, and M.~Morzfeld.
\newblock Performance bounds for particle filters using the optimal proposal.
\newblock \emph{Mon. Weather Rev.}, 143:\penalty0 4750--4761, 2015.

\bibitem[Trevisan and Uboldi(2004)]{trevisan2004assimilation}
Anna Trevisan and Francesco Uboldi.
\newblock Assimilation of standard and targeted observations within the
  unstable subspace of the observation-analysis-forecast cycle system.
\newblock \emph{J. Atmos. Sci.}, 61\penalty0 (1):\penalty0 103--113, 2004.

\bibitem[Ueckermann et~al.(2013)Ueckermann, Lermusiaux, and
  Sapsis]{ueckermann2013numerical}
M.~Ueckermann, P.~Lermusiaux, and T.~Sapsis.
\newblock Numerical schemes for dynamically orthogonal equations of stochastic
  fluid and ocean flows.
\newblock \emph{J. Comput. Phys.}, 233:\penalty0 272--294, 2013.

\bibitem[Vallis(2006)]{Vallis}
G.~Vallis.
\newblock \emph{Atmospheric and oceanic fluid dynamics: fundamentals and
  large-scale circulation}.
\newblock Cambridge University Press, 2006.

\bibitem[Weaver and Courtier(2001)]{Weaver-Courtier01}
A.~Weaver and P.~Courtier.
\newblock Correlation modelling on the sphere using a generalized diffusion
  equation.
\newblock \emph{Quart. J. Roy. Meteor. Soc.}, 127:\penalty0 1815--846, 2001.

\end{thebibliography}


\appendix

\section{Quadratic variation}

\label{QuadVar}
The quadratic co-variation process denoted $\langle\XX,\YY\rangle _t$, 
is defined as the limit in probability over a partition $\{t_1,\ldots,t_n\}$ of $[0,t]$ with $t_1<t_2<\cdots<t_n$, and a partition spacing $\delta t_i= t_i -t_{i-1}$, noted as $|\delta t|_n= \max\limits_{i} \delta t_i$ and such that $|\delta t|_n\to 0$ when $n\to\infty$:
\begin{equation}
\nonumber
\langle\XX,\YY\rangle _t = \lim^P_{|\delta t|_n\rightarrow 0} \sum_{i=0}^{n-1} \bigl(\XX(t_{i+1}) - \XX(t_i)\bigr) \bigl(\YY(t_{i+1}) - \YY(t_i)\bigr)\transp.
\end{equation}
For Brownian motions, it follows $\langle B,B\rangle _t =t$, $\langle B,h\rangle _t=\langle h,B\rangle _t = \langle h,h\rangle _t=0$, where $h$ is a deterministic function (or a random time-differentiable function) and $B$ a scalar Brownian motion. The quadratic co-variation of the uncertainty component reads
\begin{align}
\left\langle \int_0^t   \left(\bsigma (\xx,t) \dif\B_t\right)^i,
\int_0^t \left(\bsigma(\yy,t)\dif \B_t\right )^j \right\rangle 
& = 
\int_0^t  \sum_k\int_{\Omega} \breve \sigma^{ik}(\xx,\zz,s) 
\;\breve \sigma^{jk} (\yy,\zz,s)\dif s\dif \zz
,
\nonumber\\
&\defin  
\int_0^t a^{ij}(\xx,\yy,s) \dif s  
.
\label{quadvar}
\end{align}
Its time derivative corresponds to the spatial covariance tensor. The diagonal of this tensor, denoted the variance tensor, corresponds to $\xx = \yy$. For isotropic random fields, $\breve \bsigma \left(\xx , \zz \right)=\breve \bsigma \left(\|\xx - \zz\|_2 \right)$, the quadratic variation is a constant diagonal matrix.

\section{Link between the material derivative $\rm D_t$ and the operator $\Dt_t$}
\label{link Material-Der}

Let us assume:
\begin{align}
\Dt_t \varTheta &= f \dif t  + \mbs h \transp \dif \B_t.
\label{explicit expression material deriv}
\end{align}
By definition of $\Dt_t$ (Eq. \eqref{Mder}),
\begin{align}
\Dt_t \varTheta 
&=
 \dif_t \varTheta 
 + \left( \w^* \dif t + \bsigma \dif \B_t \right ) \bcdot \bnabla \varTheta
 - \tfrac{1}{2} \bnabla \bcdot \left( \mbs a \bnabla \varTheta \right ) \dif t.
 \label{appendix expression DD}
\end{align}
It yields:
\begin{align}
 \dif_t \varTheta
&=
 \left ( 
 f
 - \w^*   \bcdot \bnabla \varTheta 
 + \tfrac{1}{2} \bnabla \bcdot \left( \mbs a \bnabla \varTheta \right ) 
 \right ) \dif t
 +  \mbs h \transp \dif \B_t 
 - \left(  \bsigma \dif \B_t \right ) \bcdot \bnabla \varTheta.
\end{align}
Denoting $\mbs H_\varTheta$ the Hessian of the function $\varTheta$, we have:
\begin{align}
 \dif_t \bnabla \varTheta
&=
 \bnabla \left ( 
 f
 - \w^*   \bcdot \bnabla \varTheta
 + \tfrac{1}{2} \bnabla \bcdot \left( \mbs a \bnabla \varTheta \right ) 
 \right ) \dif t
 + \bnabla  \mbs h \transp \dif \B_t 
 - \bnabla \left(  \bsigma \dif \B_t \right ) \transp \bnabla \varTheta 
 - \mbs H_\varTheta \left(  \bsigma \dif \B_t \right ) .
 \label{gradient Theta}
\end{align}
As $\varTheta$ is a random function, its material derivative, {\rm i.e.} the differential of  $\varTheta(t,\XX_t)$,  involves the composition of two stochastic processes. Its evaluation requires the use of a generalized Ito formula, referred to as the Ito-Wentzell formula \citep[see theorem 3.3.1,][]{Kunita}. In the same way as the classical Ito formula\footnote{relevant only to express the differential of a time-differentiable function of a stochastic process.}, it incorporates quadratic variation terms related to the process $\XX_t$, but also co-variation terms between $\XX_t$ and the gradient of the random function $\varTheta$, as: 
\begin{eqnarray}
 \left (  \rm D_t \varTheta   \right ) \left(t, \XX_t \right )
& \defi &
 \dif \left ( \varTheta \left (t, \XX_t \right ) \right ),\\
&=&
 \dif_t \varTheta 
 + \dif \XX_t \bcdot \bnabla \varTheta
+ \tfrac{1}{2} \tr \left ( \dif < \XX_t, \XX_t \transp> \mbs H_\varTheta \right) 
+  \dif < \XX_t \transp , \bnabla \varTheta >
,\\
&=&
\nonumber
 \dif_t \varTheta 
 + \left( \w \dif t + \bsigma \dif \B_t \right ) \bcdot \bnabla \varTheta
+ \tfrac{1}{2} \tr \left ( \mbs a \mbs H_\varTheta \right) \dif t
+\tr \left ( \bsigma \transp \bnabla  \mbs h \transp \right )  \dif t
\\
& &
 - \sum_{k=1}^d \bsigma_{\bullet k} \transp  \bnabla \bsigma_{\bullet k}  \transp \bnabla \varTheta \dif t
 - \tr \left ( \bsigma \transp   \mbs H_\varTheta  \bsigma \right ) \dif t
,
\hspace{1cm} \text{ (using (\ref{gradient Theta}))}
\\
&=&
\nonumber
 \dif_t \varTheta
 + \left( \w \dif t + \bsigma \dif \B_t \right ) \bcdot \bnabla \varTheta
- \tfrac{1}{2} \tr \left ( \mbs a \mbs H_\varTheta \right) \dif t
\\
& &
+\tr \left ( \bsigma \transp \bnabla  \mbs h \transp \right )  \dif t
 - \left( \dv \mbs a  - \dv \bsigma \bsigma \transp \right)\bnabla \varTheta \dif t
 ,
\\
&=&
\nonumber
 \dif_t \varTheta 
 + \left( \left( \w - \tfrac{1}{2} \left(\dv \mbs a\right)\transp 
 + \bsigma (\dv \bsigma ) \transp 
 \right) \dif t 
 + \bsigma \dif \B_t \right ) \bcdot \bnabla \varTheta
\\
& &
- \tfrac{1}{2} \dv \left ( \mbs a  \bnabla \varTheta \right ) \dif t
+\tr \left ( \bsigma \transp \bnabla  \mbs h \transp \right )  \dif t
,\\
&=&
 \Dt_t \varTheta
+\tr\left ( \bsigma \transp \bnabla  \mbs h \transp \right )  \dif t.
\hspace{3cm} \text{ (by definition of $\Dt_t$)}
\end{eqnarray}
Finally, taking this Lagrangian formulation at $\XX_t=\xx$ leads to the (Eulerian) expression of the material derivative:
\begin{align}
\rm D_t \varTheta 
& \defi
\left ( \dif_t \left ( \varTheta \left ( t, \XX_t \right ) \right ) \right )_{|_{\XX_t=\xx}}
=
 \Dt_t \varTheta 
+\tr\left ( \bsigma \transp \bnabla  \mbs h \transp \right )  \dif t.
\label{appendix link D DD}
\end{align}
Conversely, assuming that the explicit expression (\ref{explicit expression material deriv}) is unknown whereas the expression of the material derivative is known:
\begin{align}
\rm D_t \varTheta &= \tilde f \dif t  + \mbs {\tilde h} \transp \dif \B_t.
\end{align}
Using the equation (\ref{appendix link D DD})
\begin{align}
\Dt_t \varTheta
& =
\rm D_t \varTheta
- \tr\left ( \bsigma \transp \bnabla  \mbs h \transp \right )  \dif t
=
\left (f - \tr\left ( \bsigma \transp \bnabla  \mbs h \transp \right )  \right )  \dif t
+ \mbs h \transp \dBt
.
\end{align}
By uniqueness of the martingale decomposition (term in $\dif t$ and term in $\dif \B_t$), we can identify $\mbs{\tilde h} = \mbs h$. Then, using again (\ref{appendix link D DD}) yields:
\begin{align}
 \Dt_t \varTheta 
=
 \rm D_t \varTheta 
- \tr\left ( \bsigma \transp \bnabla  \mbs h \transp \right )  \dif t
=
 \rm D_t \varTheta
- \tr\left ( \bsigma \transp \bnabla  \mbs{\tilde h} \transp \right )  \dif t.
\end{align}


\section{The evolution of the variance of a passive tracer}
\label{variance tracer proof}

For a passive scalar $\varTheta$, we denote $Y \defi \varTheta- \E \varTheta)$ and $Z \defi Y^2$. The goal is to find the evolution of $\vq = \E Z)$. The conservation of the tracer, says $\Dt_t \varTheta=0$, gives the evolution equation of $Y$:
\begin{eqnarray}
\dif_t Y
= -\left ( \w^{\star} \bcdot \nab \right ) Y \dif t 
+ \nab \bcdot \left ( \tfrac{1}{2}\mbs a \nab Y \right ) \dif t
- \left ( \bsigma \dBt \bcdot \nab \right ) \varTheta.
\end{eqnarray}
And, by the Ito formula,
\begin{eqnarray}
\dif_t Z &=& 2 Y \dif_t Y + \dif_t <Y,Y> ,\\
 &=&  - \w^{\star}  \bcdot  \nab Z \dif t 
 + Y \dv \left ( \mbs a \nab Y  \right ) \dif t 
 -2Y \left ( \bsigma \dBt \bcdot \nab \right ) \varTheta
 + \left(   \nab \varTheta  \right) \transp \mbs a \nab \varTheta  \dif t .
\end{eqnarray}
 Taking the expectation of this expression and using $\varTheta= \Exp(\varTheta) + Y$, yields
\begin{multline}
\partial_t \vq
=
  -   \w^{\star}  \bcdot  \nab \vq 
 + \mathbb{E} \left \{  Y \dv \left ( \mbs a \nab Y \right ) \right \} 
 +  \left(   \nab \Exp(\varTheta)   \right) \transp \mbs a \nab \Exp(\varTheta)  
 + \mathbb{E} \left \{ \left(   \nab Y  \right) \transp \mbs a \nab Y \right \}   .
\end{multline}
Expanding the second term of the right-hand side makes appear the diffusion of the variance
\begin{eqnarray}
\mathbb{E} \left \{   Y \dv \left ( \mbs a \nab Y \right ) \right \} 
 &=& \sum_{i,j} \dfi a_{ij}\mathbb{E} \left \{ Y   \dfj Y \right \} 
 + \sum_{i,j}  a_{ij} \mathbb{E} \left \{ Y  \partial_{ij}^2 Y \right \} ,\\
 &=& \sum_{i,j} \dfi a_{ij} \mathbb{E} \left \{ Y   \dfj Y \right \}  
 + \sum_{i,j}  a_{ij} \mathbb{E} \left \{ 
 \tfrac{1}{2} \partial_{ij}^2 (Y^2) 
 - \dfi Y \dfj  Y 
 \right \} ,\\
 &=& \tfrac{1}{2} \sum_{i,j} \dfi a_{ij} \dfj  \mathbb{E} \left \{ Z \right \} 
 + \tfrac{1}{2} \sum_{i,j}  a_{ij} \partial_{ij} \mathbb{E} \left \{ Z  \right \}  
 -  \mathbb{E} \left \{ \left(   \nab Y  \right) \transp \mbs a \nab Y  \right \} 
 ,\\
 &=&   \dv \left ( \tfrac{1}{2} \mbs a \nab \vq \right ) 
 - \mathbb{E} \left \{ 
 \left(   \nab Y  \right) \transp \mbs a \nab Y 
 \right \}
 .
\end{eqnarray}
Finally, the evolution law of the variance writes
\begin{eqnarray}
\partial_t \vq
 +  \w^{\star}  \bcdot  \nab \vq  
 &=&  \dv \left ( \tfrac{1}{2} a \nab \vq \right ) 
 +  \left(   \nab \Exp (\varTheta)  \right) \transp \mbs a \nab \Exp (\varTheta)  . 
\end{eqnarray}


\section{Stochastic extension of the Reynolds transport theorem}
\label{Reynolds}
 
 
 In the following, we consider a scalar function $\phi$ transported by the stochastic flow $\xx_0 \mapsto \xx = \XX_t(\xx_0)$ \eqref{particle_dX}. Its initial time value  $g$: 
\bea
\phi(\XX_t(\xx_0),t)= g(\xx_0).
\eea
We will assume that the initial function $g:\Omega\rightarrow \reel$ has bounded spatial gradients and vanishes outside the initial volume ${\cal V}(t_0)$ and on its boundary. 
The material derivative of $\phi$ is:
\bea
( \rm D_t \phi )  (t,\XX_t(\xx_0)) \defi \dif \left(  \phi (t,\XX_t(\xx_0)) \right) = \dif g(\xx_0) =0.
\eea
With equation \eqref{link DD and material deriv}, it writes in the Eulerian space:
\bea
0 = \Dt_t \phi \defi 
\dif_t \phi + 
\left ({\w}^\star\dif t + \bsigma \dif\B_t \right)\bcdot \nab \phi
 - \nab\bcdot \left ( \tfrac{1}{2} \mbs a  \nab \phi \right )\dif t 
, 
\eea
with
\bea
  \w^\star = \w   - \tfrac{1}{2} ( \nab\bcdot \mbs a)\transp 
  + \bsigma (\dv \bsigma ) \transp.
\eea
Thus,
\begin{align}
\label{SMD}
\dif_t\phi   
&=  {\cal L} \phi  \dif t  -  {\nab}\phi\,\bcdot\,\bsigma \dif\B_t, \\
{\cal L} \phi 
&= -{\nab}\phi\,\bcdot\,{\w}^\star 
+ \tfrac{1}{2} \nab\bcdot \, (\mbs a \nab \phi).
\label{defL}
\end{align}  
Denoting $J$ the Jacobian corresponding to the change of variables $\xx_0 \mapsto \xx = \XX_t(\xx_0)$,  the differential of the integral over a material volume of the product $q\phi$ is given by
\begin{align}
\dif\int_{{\cal V}(t)}
\!\!\!\! (q\phi)
(\xx,t)\dif \xx
&= 
\dif\int_{{\cal V}(0)} 
\!\! (Jq\phi) (\XX_t(\xx_0),t)\dif \xx_0
,\\
&= 
\dif\int_{\Omega} 
\!\! (Jq\phi) (\XX_t(\xx_0),t)\dif \xx_0
,\\
&= 
\dif\int_{\Omega} 
\!\! (q\phi) (\xx,t)\dif \xx
,\\
&= 
\int_{\Omega} \!\bigl(\dif_t q \phi + q \dif_t\phi  +\dif_t\langle q,\phi\rangle \bigr)(\xx,t) \dif \xx
,
\end{align}
where the second line comes from $\phi(\XX_t(\xx_0),t)=g(\xx_0)= 0 \mbox{ if }\xx_0\in\Omega\backslash {\cal V}(t_0)$
and the last line from the Ito's formula. To compute the quadratic covariation $\dif_t\langle q,\phi\rangle$, we introduce a notation for the non-differentiable part ({\rm i.e.} the integral in $\dBt$) of $\int_0^t \Dt_t q$:
\bea
\Dt_t q = f \dif t + \mbs h \transp \dBt.
\label{appendix def h for Dt q}
\eea
Together with the stochastic operator, $\Dt_t$, this relation determines the form of the time differential of $q$:
\bea 
\dif_t q = m \; \dif t + \left( - {\nab}q\transp \bsigma + \mbs h \transp \right) \dif \B_t .
\eea
Hence, from (\ref{SMD}), we have 
\begin{equation}
 \dif \int_{\Omega} q\phi
 =
 \int_{\Omega}
 \biggl [
 \dif_t q \phi 
 + q \bigl( {\cal L} \phi \dif t - {\nab}\phi\,\bcdot \, \bsigma \dif \B_t \bigr )
 - \nab \phi\transp \bsigma 
 \left( - \bsigma \transp  \nab q + \mbs h \right)
 \dif t
\biggr ] 
 .
\end{equation}
Introducing ${\cal L}^\ast$ the (formal) adjoint of the operator ${\cal L}$ in the space $L^2(\Omega)$ with Dirichlet boundary conditions, this can be written as
\begin{equation}
\int_{\Omega} \biggl [
\dif_t q + \bigl({\cal L}^\ast q -\nab\bcdot \,(\mbf a \nab q )
  + \nab \bcdot \left(  \bsigma \mbs h \right)
\bigr)\dif t 
+ \!\nab\bcdot\bigl(q \bsigma \dif\B_t\bigr) 
\biggr ]
\phi
.
\end{equation}
With the complete expression of ${\cal L}^\ast$ (the second right-hand term of \ref{defL} is self-adjoint), the condition $\phi(\xx,t) \rightarrow \car_{{\cal V}(t) /\partial{\cal V}(t)}$, where $\car$ stands for the characteristic function, leads to the following form of this differential:
\bea
\dif \int_{{\cal V}(t)} q 
  &=&
\int_{{\cal V}(t)}\! 
\biggl[
\dif_t q 
+\biggl( {\nab} \bcdot\,\bigl( q\w^\star 
\bigr) 
   + \nab \bcdot \left(  \bsigma \mbs h \right)
   \biggr)  \dif t
+ \!\nab\bcdot\bigl(q \bsigma \dif\B_t\bigr) 
\biggr]
,\\
 &=&
\int_{{\cal V}(t)}\! 
\biggl[
\Dt_t q 
  + \tr\left(  \bsigma \transp \nab \mbs h \transp \right) \dif t
  + (\dv \bsigma )\mbs h  \; \dif t
  + \nab \bcdot
  \left( \w^\star   \dif t 
    + \bsigma \dBt 
    \right )q
   \biggr]
,\\
  &=&
\int_{{\cal V}(t)}\! 
\biggl[
\rm D_t q 
  + \nab \bcdot
  \left( \w^\star   \dif t 
  + \bsigma \dBt 
  \right )q
  + (\dv \bsigma )\mbs h  \; \dif t
  \biggr]
,\\
  &=&
\int_{{\cal V}(t)}\! 
\biggl[
\rm D_t q 
  + \nab \bcdot
  \left( \w^\star   \dif t 
  + \bsigma \dBt 
  \right )q
  +
 \dif  \left <
  \int_0^t D_{t'} q ,
  \int_0^t \dv \bsigma \dif \B_{t'}
   \right >
  \biggr]
,\\
\eea
where the third line comes from the explicit link \eqref{link DD and material deriv}, between the stochastic transport operator $\Dt_t$ and the material derivative $\rm D_t$.

As a simple  example  of these formulas application, the rate of change of a passive scalar quantity within a  material volume  (i.e $\mbs h=0$) for  a divergent uncertainty random field, reads:
\bea
\dif \int_{{\cal V}(t)} q 
  &=&
\int_{{\cal V}(t)}\! 
\biggl[
\rm D_t q 
  + \nab \bcdot
  \left( \w^\star   \dif t 
  + \bsigma \dBt 
  \right )q \biggr] 
  ,\\
 &=&
\int_{{\cal V}(t)}\! 
\biggl[
\Dt_t q 
  + \nab \bcdot
  \left( \w^\star   \dif t 
    + \bsigma \dBt 
    \right )q
   \biggr],\\
   &=&
\int_{{\cal V}(t)}\! 
\biggl[
\dif_t q 
+ {\nab} \bcdot\,\biggl( q \bigl(\w   - \tfrac{1}{2} ( \nab\bcdot \mbs a)\transp 
  + \bsigma (\dv \bsigma ) \transp\bigr)
\biggr) \dif t
+ \!\nab\bcdot\bigl(q \bsigma \dif\B_t\bigr) 
\biggr]
.
\eea


\section{Stochastic Navier-Stokes model}
\label{Appendix Stochastic Navier-Stokes model}

From the conservation of linear momentum, the balance between the momentum variation and the forces can be expressed as:
\bea 
\dif \int_{{\cal V}(t)} \!\!\!\!\!\!\!\!\!{\rho\bigl(\w +\bsigma \dot{\mbs B}\bigr)}
=  \int_{{\cal V}(t)}\!\!\!\!\!\!\!\!\! \dif_t {\boldsymbol{F}}.
\eea
The left-hand term must be interpreted in a distribution sense, the small-scale velocity, $\bsigma \dot{\mbs B}$, being non-continuous. For every test function  $h\in C_0^\infty(\R_+)$, we have:
\bea
\label{eq-dist1} 
 \int_{\R_+} h(t) \dif \int_{{\cal V}(t)}\!\!\!\!\!\!\!\!\!
 \rho\w \,
 - \int_{\R_+} \frac{\dif h}{\dif t}(t) \int_{{\cal V}(t)}\!\!\!\!\!\!\!\!\!\rho\bsigma\dif \B_t
= \int_{\R_+} h(t) \int_{{\cal V}(t)}\!\!\!\!\!\!\!\!\!  \dif_t {\boldsymbol{F} } .
\eea 
%
%
%
Both sides of this equation must have the same structure, and the forces can be written as:
\bea
\label{eq-dist2} 
\int _{\R_+}h(t) \int_{{\cal V}(t)}\!\!\!\!\!\!\!\!\! \dif_t  \boldsymbol{F}
=
- \int_{\R_+} \frac{\dif h}{\dif t}(t) \int_{{\cal V}(t)}\!\!\!\!\!\!\!\!\!\rho\bsigma\dif \B_t 
+
 \int _{\R_+}h(t) \int_{{\cal V}(t)}\!\!\!\!\!\!\!
 \left(\mbs \eta \dif t + \mbs \theta \dBt \right)
.
\eea
The right-hand first term must compensate the white-noise distributional differentiation of (\ref{eq-dist1}), whereas the last term of (\ref{eq-dist2}) provides the structure of the forces under location uncertainty.
The forces are due to the gravitation potential  $\Phi_a$ within the absolute frame, pressure and friction forces, $ \dif_t  {\cal F} (\w,\bsigma) $. A direct stochastic extension of the deterministic forces expression reads:  
\begin{equation}
\label{sto-stress}
\int_{{\cal V}(t)}\!\!\!\!\!\!\!
 \left(\mbs \eta \dif t + \mbs \theta \dBt \right)
 =
 \int_{{\cal V}(t)}
 \left(
 \rho \nab \Phi_a \dif t- \nab (p\dif t +\dif_t p_{\sigma}) + \dif_t  {\cal F} (\w,\bsigma) 
 \right).
\end{equation}
The pressure term $p$ denotes the continuous contribution of the pressure. The other term, $
\dot{ p_{\sigma}}
$, is a zero-mean non-continuous stochastic process (the white noise part of the pressure). It describes the pressure fluctuations due to the random velocity component.  
Note that the gravity force is continuous in time, whereas the friction force applies both on the deterministic and stochastic velocity components.
For a fixed observer in a rotating frame, the  rate of change of the fluid velocity incorporates (considering the rotation, $\mbs f$,  constant in time) the centripetal acceleration and the Coriolis acceleration as additional terms. The centrifugal force is included  within an effective gravity, $\mbs  g =-\nab \Phi$. The Coriolis term applies both to the large-scale component of the velocity and to the random small-scale field. 

The transport equation applied to the linear momentum gives:
\bea
\dif\!\!\int_{{\cal V}(t)}\!\!\rho \w
&=&
\int_{{\cal V}(t)}\!\! \Dt_t \left( \rho \w \right) 
+ \rho \w\nab\bcdot \w^\star \dif t.
\label{appendix th transport for momentum}
\eea 

With $\Dt_t$ given by \eqref{Mder}, the equation \eqref{appendix th transport for momentum} can be expressed in terms of $\rho$, $\w$ and $\dif_t(\rho \w)$. The large-scale velocity $\w$ is assumed to be differentiable in time,
\bea 
\label{appendix time deriv mass velocity}
\dif_t(\rho \w) = 
\dif_t\rho \w + \rho \partial_t \w \dif t.
\eea
The density time derivative, $\dif_t \rho$ uses \eqref{Mder} and the mass conservation equation:
\begin{align}
\label{appendix mass for momentum}
&\Dt_t \rho + \rho \nab\bcdot \w^\star =0
.
\end{align}
From equations \eqref{appendix th transport for momentum}, \eqref{Mder}, \eqref{appendix time deriv mass velocity} and \eqref{appendix mass for momentum}, the variation of the large-scale linear momentum reads:
\bea 
\dif\!\!\int_{{\cal V}(t)}\!\!\rho w_{i} 
&=&
\int_{{\cal V}(t)}\!\!\Bigl( 
\rho \bigl(\partial_t  w_{i} \dif t
+ \rho \left( \w^*  \dif t + \bsigma \dif \B_t \right ) \bcdot \nab  w_{i} 
- \tfrac{1}{2}\nab \bcdot \left( \rho \mbs a  \nab w_i \right) \dif t 
\Bigr).
\label{transport-momentum}
\eea 

From the balance between the forces \eqref{sto-stress} and the momentum variation (\ref{transport-momentum}), the expression of the flow dynamics for an observer in an uniformly rotating coordinate frame is then obtained by considering the slow temporal bounded variation terms and the Brownian terms.


\end{document}